\documentclass[12pt]{article}
\usepackage{amsmath, amsfonts,amsbsy,amssymb}
\usepackage{graphicx,psfrag,epsf}
\usepackage{natbib}
\usepackage{bbm}
\usepackage{color}
\usepackage{enumitem}
\usepackage{booktabs} 

\newtheorem{theorem}{Theorem}[section]

\newtheorem{lemma}{Lemma}[section]
\newtheorem{corollary}{Corollary}[section]

\DeclareMathOperator{\E}{\mathbb{E}}
\DeclareMathOperator{\Cov}{Cov}
\DeclareMathOperator{\Bias}{Bias}
\DeclareMathOperator{\V}{\mathbb{V}}
\DeclareMathOperator{\mydiag}{diag}
\DeclareMathOperator{\supp}{supp}
\DeclareMathOperator{\AMISE}{AMISE}

\newcommand{\blind}{0}

\begin{document}

\def\spacingset#1{\renewcommand{\baselinestretch}%
{#1}\small\normalsize} \spacingset{1}


\if0\blind
{
  \title{\bf Inference for Sparse and Dense Functional Data with Covariate Adjustments}
  \author{Dominik Liebl\\
    Institute for Financial Economics and Statistics, University of Bonn
    }
  \maketitle
} \fi

\if1\blind
{
  \bigskip
  \bigskip
  \bigskip
  \begin{center}
    {\LARGE\bf Inference for Functional Data with Covariate Adjustments: From Sparse to Dense}
\end{center}
  \medskip
} \fi

\bigskip
\begin{abstract}
We consider inference for the mean and covariance functions of covariate adjusted functional data using Local Linear Kernel (LLK) estimators. By means of a double asymptotic, we differentiate between sparse and dense covariate adjusted functional data -- depending on the relative order of $m$ (the discretization points per function) and $n$ (the number of functions).
Our simulation results demonstrate that the existing asymptotic normality results can lead to severely misleading inferences in finite samples. We explain this phenomenon based on our theoretical results and propose finite-sample corrections which provide practically useful approximations for inference in sparse and dense data scenarios. The relevance of our theoretical results is showcased using a real-data application.
\end{abstract}

\noindent%
{\it Keywords:}
functional data analysis,
local linear kernel estimation,
asymptotic normality,
multiple bandwidth selection,
finite-sample correction
\vfill

\newpage
\spacingset{1.2}

\section{Introduction}\label{sec:intro}
This work considers the case of independently identically distributed (iid) covariate adjusted functional data $X_i(.,Z_i)\in L^2([0,1])$, $i=1,\dots,n$, with random covariate $Z_i\in[0,1]\subset\mathbb{R}$. As typically for longitudinal data, the single functions are only observed at error-prone measurements sampled at a certain number of random locations. That is, each unobserved random function $X_i(.,Z_i)$ is observed at $m$ data points data points $(Y_{ij},U_{ij})\in\mathbb{R}^2$ with
\begin{align}\label{eq:basic_1}
  Y_{ij}&=X_i(U_{ij},Z_i)+\epsilon_{ij},\quad j=1,\dots,m,\quad i=1,\dots,n,
\end{align}
where $\epsilon_{ij}\in\mathbb{R}$ is an iid error term with mean zero and $\V(\epsilon_{ij})=\sigma^2_\epsilon<\infty$ independent from $X_i$, $U_{ij}$, and $Z_i$.

We derive inferential results for the LLK estimators of the mean function $\mu(u,z)=\E(X_i(u,z))$ and the covariance function $\gamma(u_1,u_2,z)=\Cov(X_i(u_1,z),X_i(u_2,z))$. So far the only other existing asymptotic normality results in this context are those of \cite{jiang2010covariate}, who consider the case of sparse covariate adjusted functional data, where sparse refers to the asymptotic scenario with $m$ being bounded while $n\to\infty$ (i.e., a finite-$m$ asymptotic). However, as shown in our simulation study, the asymptotic variance expressions derived in \cite{jiang2010covariate} tend to severely underestimate the actual variances in finite samples. This can result in false inferences (size-distortion) in finite samples.

We are able to explain this finding based on our asymptotic normality results. The finite-$m$ asymptotic considered by \cite{jiang2010covariate} neglects an additional functional-data-specific variance term which is typically not negligible in practice. In contrast to \cite{jiang2010covariate}, we consider sparse and dense functional data depending on the relative order of $m$ and $n$. This approach is related to the work of \cite{zhang2016sparse}, who, however, consider classical functional data \emph{without} covariate adjustments\footnote{Our results are based on the  author's PhD thesis \citep{L13_thesis} and were developed independently from the work of \cite{zhang2016sparse}.}.

Additionally, we derive the explicit optimal multiple bandwidth expressions for the case of sparse and dense covariate adjusted functional data. For dense functional data, this leads to rather unconventional bandwidth expressions with different convergence rates for the bandwidths in $U$- and $Z$-direction. Effectively, this imposes a necessary under-smoothing in $U$-direction, which guarantees that the $U$-related bias and variance components become negligible in comparison to the $Z$-related bias and variance components.

Our third contribution is concerned with finite-samples. The differentiation between sparse and dense functional data is based on pure theoretical considerations. In practice, however, it is usually impossible to differentiate between these two asymptotic data scenarios. Therefore, we contribute finite-sample corrections that allow for robust inferences with sparse and dense functional data.

Generally, there are many different concepts of sparsity and we refer to \cite{AV16} for a comprehensive overview. Throughout this paper, we use the terms sparse and dense in order to differentiate between the following two asymptotic scenarios:
\begin{center}
sparse: $m/n^{1/5}\to 0$\quad and\quad dense: $m/n^{1/5}\to\infty$,
\end{center}
where the value $1/5$ of the exponent is determined by our theory. The sparse asymptotic scenario approximates cases where $m$ is relatively small in comparison to $n^{1/5}$, i.e., very small in comparison to $n$, and includes the finite-$m$ asymptotic of \cite{jiang2010covariate}. The dense asymptotic scenario approximates cases where $m$ is relatively large in comparison to $n^{1/5}$; however, not necessarily large in comparison to $n$. The terminology of sparse and dense asymptotic scenarios refers to the work of \cite{zhang2016sparse}.

The term dense, however, must be used with caution as it has the here misleading connotation of many data points $m$, which not necessarily applies to this asymptotic scenario\footnote{Indeed, \cite{zhang2016sparse} use the term ultra-dense which has a potentially even more misleading connotation.}, since it includes cases where $m$ is relatively \emph{small} in comparison to $n$, i.e., scenarios with possibly not so many data points $m$. Indeed, for large $m$ it is usually advantageous to pre-smooth the single functions and to neglect the pre-smoothing error \citep{ZC07}. In this paper, we focus on cases where the pre-smoothing approach cannot be applied due to a too small $m$. 

The literature on covariate adjusted functional data was initiated by the work of \cite{cardot2007conditional}, who considers functional principal component analysis for dense functional data, but does not provide inferential results. \cite{jiang2010covariate} focus on the case of sparse functional data. \cite{LSB14} consider a copula-based model and \cite{ZW15} propose an iterative algorithm for computing functional principal components, though neither provides inferential results for the covariate adjusted mean and covariance functions.
For the case without covariate adjustments there are several papers considering inference. \cite{ZC07} and \cite{HK07} consider inference in the pre-smoothing context for dense functional data. \cite{FMV07}, \cite{FKV10}, and \cite{RAVV16} consider inference in functional nonparametric regression. \cite{Benko2009a} develop bootstrap procedures for the case of dense functional data. \cite{CYT12} derive simultaneous confidence bands in the case of dense functional data. \cite{GK12} consider an $L^2$-based test statistic and address computational issues in finite samples and \cite{HKR13} focus on the case of dependent functional data within the same framework. Although related, the case without covariate adjustments is fundamentally different from our case, since the presence of a covariate affects the involved bandwidth selection problem in a nontrivial manner. Readers with a general interest in functional data analysis are referred to the textbooks of \cite{RamsayfdaBook2005}, \cite{Ferraty2006}, \cite{horvath2012inference}, \cite{HR15_book}, and \cite{KR17_book}. Recent surveys of methodological advances in functional data analyses can be found in \cite{C14}, \cite{GV16}, and \cite{WCM2016}.

The rest of this paper is structured as following. The next section introduces the considered regression models and LLK estimators. Section \ref{sec:AssRes} presents our assumptions and asymptotic results.  Our simulation study is in Section \ref{sec:Sim}. Section \ref{sec:ROT} introduces rule-of-thumb approximations to our theoretical bandwidth expressions and practical plug-in estimates for the unknown bias and variance components. Section \ref{sec:App} contains our real data application. All proofs can be found in the appendix.

\section{Nonparametric regression models and estimators}\label{sec:ModEst}
Let $X^c_i$ denote the centered random function $X^c_i(U_{ij},Z_i)=X_i(U_{ij},Z_i)-\E(X_i(U_{ij},Z_i)|\mathbf{U},\mathbf{Z})$, where $\mathbf{U}=(U_{11},\dots,U_{nm})^\top$ and $\mathbf{Z}=(Z_1,\dots,Z_n)^\top$. Model \eqref{eq:basic_1} can be written as a nonparametric regression model with the bivariate mean function
$\mu(U_{ij},Z_i)=\E(X_i(U_{ij},Z_i)|\mathbf{U},\mathbf{Z})$
as the regression function,
\begin{align}\label{Eq2}
  Y_{ij}=\mu(U_{ij},Z_i)+X^c_i(U_{ij},Z_i)+\epsilon_{ij},\quad i=1,\dots,n,\quad j=1,\dots,m,
\end{align}
where $X^c_i(.,z)\in L^2[0,1]$ is an iid centered random function, $U_{ij}\in[0,1]$ and $Z_i\in[0,1]$ are iid random predictors, and $\epsilon_{ij}\in\mathbb{R}$ is an iid random error term independent from $X_i^c$, $U_{ij}$ and $Z_i$. Note that Model \eqref{Eq2} has a rather unusual composed error term $X^c_i(U_{ij},Z_i)+\epsilon_{ij}$ consisting of a function- and a scalar-valued component. This structure of the error term leads to an additional functional-data-specific variance term.

Likewise to Model \eqref{Eq2} we can define the following nonparametric regression model with the trivariate covariance function $\gamma(U_{ij},U_{ik},Z_i)=\Cov(X_i(U_{ij},Z_i),X_i(U_{ik},Z_i)|\mathbf{U},\mathbf{Z})$ as the regression function:
\begin{align}
  C_{ijk}&=\gamma(U_{ij},U_{ik},Z_i)+\tilde{X}^c_i(U_{ij},U_{ik},Z_i)+\varepsilon_{ijk},\quad i=1,\dots,n,\quad j\neq k\in\{1,\dots,m\},\label{Eq3}
\end{align}
where the raw-covariances $C_{ijk}$, the centered random function $\tilde{X}^c_i$, and the scalar-valued error term $\varepsilon_{ijk}$ are defined as
\begin{align}
  C_{ijk}&=(Y_{ij}-\mu(U_{ij},Z_i))(Y_{ik}-\mu(U_{ik},Z_i)),\label{RawCov}\\
  \tilde{X}^c_i(U_{ij},U_{ik},Z_i)&=X^c_i(U_{ij},Z_i)\,X^c_i(U_{ik},Z_i)-\gamma(U_{ij},U_{ik},Z_i),\;\text{and}\notag\\
\varepsilon_{ijk}&=X^c_i(U_{ij},Z_i)\epsilon_{ik}+X^c_i(U_{ik},Z_i)\epsilon_{ij}+\epsilon_{ij}\epsilon_{ik}.\notag
\end{align}

In contrast to $\epsilon_{ij}$, the scalar error term $\varepsilon_{ijk}$ is heteroscedastic with
$\V(\varepsilon_{ijk})=\sigma^2_\varepsilon(u_1,u_2,z)$, where $\sigma^2_\varepsilon(u_1,u_2,z)=\gamma(u_1,u_1,z)\,\sigma^2_\epsilon+\gamma(u_2,u_2,z)\sigma^2_\epsilon+\sigma^4_\epsilon$. Note that $\E(\varepsilon_{ijk})\neq 0$ for all $j=k$, therefore all raw covariance points $C_{ijk}$ with $j=k$ need to be excluded \citep[see also][]{Yao2005}. Correspondingly, the number of raw covariance points for each $i$ is $M=m^2-m$, which makes it necessary that $m\geq 2$. As in Model \eqref{Eq2}, the error term of Model \eqref{Eq3}, $\tilde X^c_i(U_{ij},U_{ik},Z_i)+\varepsilon_{ijk}$, consists of a function- and a scalar-valued component.

We estimate the mean function $\mu(u,z)$ using the LLK estimator $\hat{\mu}(u,z;h_{\mu,U},h_{\mu,Z})$ defined as the following locally weighted least squares estimator \citep[see, e.g.,][]{ruppert1994}:
\begin{align}
\hat{\mu}(u,z;h_{\mu,U},h_{\mu,Z})=e_1^{\top}\left([\mathbf{1},\mathbf{U}_u,\mathbf{Z}_z]^{\top}\mathbf{W}_{\mu,uz}[\mathbf{1},\mathbf{U}_u,\mathbf{Z}_z]\right)^{-1}[\mathbf{1},\mathbf{U}_{u},\mathbf{Z}_z]^{\top}\mathbf{W}_{\mu,uz}\mathbf{Y},\label{Estimator_mu}
\end{align}
where the vector $e_1=(1,0,0)^{\top}$ selects the estimated intercept parameter and $[\mathbf{1},\mathbf{U}_u,\mathbf{Z}_z]$ is a partitioned $nm\times 3$ dimensional data matrix with typical rows $(1,U_{ij}-u,Z_i-z)$. The $nm\times nm$ dimensional diagonal weighting matrix $\mathbf{W}_{\mu,uz}$ holds the bivariate multiplicative kernel weights
$K_{\mu,h_{\mu,U},h_{\mu,Z}}(U_{ij}-u,Z_i-z)=h^{-1}_{\mu,U}\,\kappa(h^{-1}_{\mu,U}(U_{ij}-u))\; h^{-1}_{\mu,Z}\,\kappa(h^{-1}_{\mu,Z}(Z_i-z)),$
where $\kappa$ is a usual second-order kernel such as, e.g., the Epanechnikov or the Gaussian kernel. The usual kernel constants are denoted by $\nu_{2}(K_\mu)=\left(\nu_{2}(\kappa)\right)^2$, with $\nu_{2}(\kappa)=\int u^2\kappa(u)du$, and $R(K_\mu)=R(\kappa)^2$, with $R(\kappa)=\int\kappa(u)^2du$. All vectors and matrices are filled in correspondence with the $nm$ dimensional vector $\mathbf{Y}=(Y_{11},Y_{12},\dots,Y_{n,m-1},Y_{n,m})^{\top}$.

The LLK estimator for the covariance function $\gamma(u_1,u_2,z)$ is defined correspondingly as
\begin{align}
\begin{split}
&\hat{\gamma}(u_1,u_2,z;h_{\gamma,U},h_{\gamma,Z})=\\
&=e_1^{\top}\left([\mathbf{1},\mathbf{U}_{u_{1}},\mathbf{U}_{u_{2}},\mathbf{Z}_z]^{\top}\mathbf{W}_{\gamma,u_{1}u_{2}z}[\mathbf{1},\mathbf{U}_{u_{1}},\mathbf{U}_{u_{2}},\mathbf{Z}_z]\right)^{-1}[\mathbf{1},\mathbf{U}_{u_{1}},\mathbf{U}_{u_{2}},\mathbf{Z}_z]^{\top}\mathbf{W}_{\gamma,u_{1}u_{2}z}\mathbf{C},\label{Estimator_gamma}
\end{split}
\end{align}
where $e_1=(1,0,0,0)^{\top}$ and $[\mathbf{1},\mathbf{U}_{u_{1}},\mathbf{U}_{u_{2}},\mathbf{Z}_z]$ is a $nM\times 4$ dimensional data matrix with typical rows $(1,U_{ij}-u_{1},U_{ik}-u_{2},Z_i-z)$. The $nM\times nM$ dimensional diagonal weighting matrix $\mathbf{W}_{\gamma,u_{1}u_{2}z}$ holds the trivariate multiplicative kernel weights $K_{\gamma,h_{\gamma,U},h_{\gamma,Z}}(U_{ij}-u_1,U_{ik}-u_2,Z_i-z)=h^{-1}_{\gamma,U}\,\kappa(h^{-1}_{\gamma,U}(U_{ij}-u_1))\; h^{-1}_{\gamma,U}\,\kappa(h^{-1}_{\gamma,U}(U_{ik}-u_2))\;h^{-1}_{\gamma,Z}\,\kappa(h^{-1}_{\gamma,Z}(Z_i-z))$, where $\kappa$ is as defined above, with kernel constants are $\nu_{2}(K_\gamma)=\left(\nu_{2}(\kappa)\right)^3$ and $R(K_\gamma)=R(\kappa)^3$. All vectors and matrices are filled in correspondence with the $nM$ dimensional vector $\mathbf{C}=(C_{112},C_{113},\dots,C_{n,m,m-2},C_{n,m,m-1})^{\top}$, where $$\hat{C}_{ijk}=(Y_{ij}-\hat{\mu}(U_{ij},Z_i;h_{\mu,U},h_{\mu,Z}))(Y_{ik}-\hat{\mu}(U_{ik},Z_i;h_{\mu,U},h_{\mu,Z})).$$

\section{Theoretical results}\label{sec:AssRes}
Before we present our asymptotic results, we list our additional assumptions which are equivalent to those in \cite{ruppert1994} with some straight forward adjustments to our functional data context.

\bigskip
\noindent\textbf{A-AS} (Asymptotic Scenario)
$nm\to\infty$, where $m=m_n\geq 2$ such that $m_n\asymp n^\theta$ with $0\leq\theta<\infty$. Here, $m_n\asymp n^\theta$ denotes that the two sequences $m_n$ and $n^\theta$ are asymptotically equivalent, i.e., that $0<\lim_{n\rightarrow\infty}(m_n/ n^\theta)<\infty$.

\bigskip
\noindent\textbf{A-RD} (Random Design)
The triple $(Y_{ij}, U_{ij}, Z_i)\in\mathbb{R}\times[0,1]^2$ has the same distribution as $(Y,U,Z)$ with pdf~$f_{YUZ}$, where $f_{YUZ}(y,u,z)>0$ for all $(y,u,z)\in\mathbb{R}\times[0,1]^2$ and zero else.
Equivalently, $(C_{ijk}, U_{ij}, U_{ik}, Z_i)\in\mathbb{R}\times[0,1]^3$ has the same distribution as $(C,U,U',Z)$ with pdf~$f_{CUUZ}$, where $f_{CUUZ}(c,u,u',z)>0$ for all $(c,u,u',z)\in\mathbb{R}\times[0,1]^3$ and zero else.

\bigskip
\noindent\textbf{A-SK} (Smoothness \& Kernel)
The pdfs $f_{YUZ}$ and $f_{CUUZ}$ and their marginals are continuously differentiable. All second-order derivatives of $\mu$  and $\gamma$ are continuous.
The multiplicative kernel functions $K_\mu$ and $K_\gamma$ are products of second-order kernel functions $\kappa$.

\bigskip
\noindent\textbf{A-MO} (Moments)
$\E((X_i(u,z))^4)<\infty$ for all $(u,z)$ and $\E(\epsilon_{ij}^2)<\infty$.

\bigskip
\noindent\textbf{A-BW} (Bandwidths)
$h_{\mu,U},h_{\mu,Z}\to 0$ and
$(nm)h_{\mu,U}h_{\mu,Z}\to\infty$ as $nm\to\infty$.
$h_{\mu,U},h_{\mu,Z}\to 0$ and
$(nM)h_{\mu,U}^2h_{\mu,Z}\to\infty$ as $nM\to\infty$.

\paragraph{Remark} Assumption A-AS is a simplified version of the asymptotic setup of \cite{zhang2016sparse}. The case $\theta=0$ implies that $m$ is bounded, which corresponds to the finite-$m$ asymptotic as considered by \cite{jiang2010covariate}. For $0<\theta<\infty$ we can consider also sparse and dense functional data. As typically done in multivariate nonparametric regressions, we focus on the case of bounded random regressors $U_{ij}$ and $Z_i$. The case of unbounded regressors is beyond the scope of this paper, but may be adapted from \cite{H2008}, who considers, however, a much simplex context.

The following two Theorems \ref{Bias_and_Variance_mu} and \ref{Bias_and_Variance_gamma} build the basis of our theoretical results.
\begin{theorem}[Bias and Variance of $\hat{\mu}$]\label{Bias_and_Variance_mu}
Let $(u,z)$ be an interior point of $[0,1]^2$. Under our setup the conditional asymptotic bias and variance of the LLK estimator in Eq.~\eqref{Estimator_mu} are then given by
\begin{align*}
      &\text{\textit{(i)}} \Bias\left\{\hat{\mu}(u,z;h_{\mu,U},h_{\mu,Z})|\mathbf{U},\mathbf{Z}\right\}=B_\mu(u,z)\left(1+o_p(1)\right)\text{ with}\hspace{3cm}\\
      &B_\mu(u,z)=\frac{1}{2}\;\nu_{2}(K_\mu)\,\left(h^2_{\mu,U}\,\mu^{(2,0)}(u,z)+h^2_{\mu,Z}\,\mu^{(0,2)}(u,z)\right),\\
      &\text{where }\mu^{(k,l)}(u,z)=(\partial^{k+l}/(\partial u^k\partial
z^l))\mu(u,z).\\[2ex]
      &\text{\textit{(ii)} }
      \V\left\{\hat{\mu}(u,z;h_{\mu,U},h_{\mu,Z})|\mathbf{U},\mathbf{Z}\right\}=\left(V_\mu^{I}(u,z)+V_\mu^{II}(u,z)\right)\left(1+o_p(1)\right)\text{
       with}\\
      &V_\mu^{I}(u,z)=(nm)^{-1}\left[h_{\mu,U}^{-1}h_{\mu,Z}^{-1}\,R(K_\mu)\frac{\gamma(u,u,z)+\sigma^2_\epsilon}{f_{UZ}(u,z)}\,\right]\text{
       and}\\
      &
      V_\mu^{II}(u,z)=n^{-1}\left[\left(\frac{m-1}{m}\right)h_{\mu,Z}^{-1}\,R(\kappa)\frac{\gamma(u,u,z)}{f_Z(z)}\right].
  \end{align*}
\end{theorem}

\begin{theorem}[Bias and Variance of $\hat{\gamma}$]\label{Bias_and_Variance_gamma}
Let $(u_1,u_2,z)$ be an interior point of $[0,1]^3$. Under our setup the conditional asymptotic bias and variance of the LLK estimator in Eq.~\eqref{Estimator_gamma} are then given by
\begin{align*}
&\text{\textit{(i)}}\Bias\left\{\hat{\gamma}(u_1,u_2,z;h_{\gamma,U},h_{\gamma,Z})|\mathbf{U},\mathbf{Z}\right\}=B_\gamma(u_1,u_2,z)\left(1+o_p(1)\right)\text{ with}\\
&B_\gamma(u_1,u_2,z)=\frac{1}{2}\nu_{2}(K_\gamma)\left(h^2_{\gamma,U}\left(\gamma^{(2,0,0)}(u_1,u_2,z)+\gamma^{(0,2,0)}(u_1,u_2,z)\right)+h^2_{\gamma,Z}\gamma^{(0,0,2)}(u_1,u_2,z)\right),\\
&\text{where }\gamma^{(k,l,m)}(u_1,u_2,z)=(\partial^{k+l+m}/(\partial u_1^k\,\partial u_2^l\,\partial z^m))\gamma(u_1,u_2,z).\\[2ex]
&\text{\textit{(ii)} } \V\left\{\hat{\gamma}(u_1,u_2,z;h_{\gamma,U},h_{\gamma,Z})|\mathbf{U},\mathbf{Z}\right\}=\left(V_\gamma^{I}(u_1,u_2,z)+V_\gamma^{II}(u_1,u_2,z)\right)\left(1+o_p(1)\right)\text{with}\\
&V_\gamma^{I}(u_1,u_2,z)=(nM)^{-1}\left[h_{\gamma,U}^{-2}h_{\gamma,Z}^{-1}\,R(K_\gamma)\frac{\tilde\gamma((u_1,u_2),(u_1,u_2),z)+\sigma^2_\varepsilon(u_1,u_2,z)}{f_{UUZ}(u_1,u_2,z)}\,\right]\text{and}\\
&V_\gamma^{II}(u_1,u_2,z)=n^{-1}\left[\left(\frac{M-1}{M}\right)h_{\gamma,Z}^{-1}\,R(\kappa)\frac{\tilde\gamma((u_1,u_2),(u_1,u_2),z)}{f_{Z}(z)}\right],
\end{align*}
\begin{equation*}
\begin{array}{rcl}
\text{where}\quad\tilde{\gamma}((u_1,u_2),(u_1,u_2),z)&=&\Cov(\tilde{X}^c_i(u_1,u_2,z),\tilde{X}^c_i(u_1,u_2,z))\quad\text{and}\\
\sigma^2_\varepsilon(u_1,u_2,z)&=&\gamma(u_1,u_1,z)\,\sigma^2_\epsilon+\gamma(u_2,u_2,z)\sigma^2_\epsilon+\sigma^4_\epsilon.
\end{array}
\end{equation*}
\end{theorem}

The bias expressions in Theorems \ref{Bias_and_Variance_mu} and \ref{Bias_and_Variance_gamma} correspond to the classical bias results \citep[see, e.g.,][]{ruppert1994}. The first variance terms $V_\mu^{I}(u,z)$ and $V_\gamma^{I}(u_1,u_2,z)$ are equivalent to those in Theorems 3.2 and 3.4 of \cite{jiang2010covariate} who consider the LLK estimators $\hat\mu(u,z)$ and $\hat\gamma(u,z)$ under the finite-$m$ asymptotic. The second functional-data-specific variance terms, $V_\mu^{II}(u,z)$ and $V_\gamma^{II}(u_1,u_2,z)$, are negligible under such a finite-$m$ asymptotic, but generally not negligible when considering a double asymptotic where both $m\to\infty$ and $n\to\infty$.

Whether the first variance terms, $V_\mu^{I}$ and $V_\gamma^{I}$, or the second, functional-data-specific variance terms, $V_\mu^{II}$ and $V_\gamma^{II}$, are the leading variance terms depends on the bandwidth choices and on the relative order of $m$ and $n$, i.e., on the value of $\theta$ in $m\asymp n^{\theta}$. In order to determine the decisive $\theta$ value we postulate optimal bandwidth choices determined from minimizing the usual Asymptotic Mean Integrated Squared Error (AMISE) criteria,
{\small\begin{equation*}
\begin{array}{l}
  \AMISE_{\hat{\mu}}=\int\big([\Bias\{\hat{\mu}(u,z)|\mathbf{U},\mathbf{Z}\}]^2+\V\{\hat{\mu}(u,z)|\mathbf{U},\mathbf{Z}\}\big)f_{UZ}(u,z)\,d(u,z)\quad\text{and}\\[2ex]
  \AMISE_{\hat{\gamma}}
=\int\big([\Bias\{\hat{\gamma}(u_1,u_2,z)|\mathbf{U},\mathbf{Z}\}]^2+\V\{\hat{\gamma}(u_1,u_2,z)|\mathbf{U},\mathbf{Z}\}\big)f_{UUZ}(u_1,u_2,z)\,d(u_1,u_2,z).\\[-1mm]
\end{array}
\end{equation*}}

In anticipation of some of our results: Under AMISE optimal bandwidth choices, the discriminating $\theta$-threshold is given by $\theta=1/5$. That is, if $m/n^{1/5}\to 0$ and $\sqrt{M}/n^{1/5}\to 0$, the first variance terms $V_\mu^{I}$ and $V_\gamma^{I}$ are the leading variance terms. This asymptotic scenario comprises situations where $m$ and $\sqrt{M}$ are eventually small in comparison to $n^{1/5}$, i.e., \emph{very} small in comparison to $n$. Following \cite{zhang2016sparse}, we refer to this asymptotic scenario as sparse covariate adjusted functional data.

If, however, $m/n^{1/5}\to\infty$ and $\sqrt{M}/n^{1/5}\to\infty$, then the functional-data-specific variance terms $V_\mu^{II}$ and $V_\gamma^{II}$ are the leading variance terms. This asymptotic scenario comprises quite general situations where $m$ and $\sqrt{M}$ are eventually large in comparison to  $n^{1/5}$, but not necessarily large in comparison to $n$. Following \cite{zhang2016sparse}, we refer to this asymptotic scenario as dense covariate adjusted functional data. However, we refer the reader to our cautionary note of the introduction as the term dense might have a misleading connotation.

Our theoretical results assume a homoscedastic variance $\sigma^2_\epsilon$ for the error term $\epsilon_{ij}$. In case of a heteroscedastic variance, one needs to replace the quantity $\sigma^2_\epsilon$ in Theorems 3.1 and 3.2 by, its heteroscedastic counterpart $\sigma^2_\epsilon(u,z)$. The unknown $\sigma^2_\epsilon(u,z)$ must then be estimated using a heteroscedasticity-consistent estimator.

\subsection{Sparse functional data}\label{FirstVarTerm}
The explicit AMISE optimal bandwidth expressions for the case of leading first variance terms, $V_\mu^{I}$ and $V_\gamma^{I}$, can be found in the following two Theorems:

\begin{theorem}[Sparse - optimal bandwidths for $\hat{\mu}$]\label{AMISE-I_opt_BDW_mu}
Let $m/n^{1/5}\to 0$ and $(u,z)$ be an interior point of $[0,1]^2$. Under our setup the AMISE optimal bandwidths for the LLK estimator in Eq.~\eqref{Estimator_mu} are then given by
\begin{align}
  h^{S}_{\mu,U}&=\left(\frac{R(K_\mu)\,Q_{\mu,1}\,\mathcal{I}_{\mu,ZZ}^{3/4}}{nm\,\left(\nu_{2}(K_\mu)\right)^2\,\left[\mathcal{I}_{\mu,UU}^{1/2}\,\mathcal{I}_{\mu,ZZ}^{1/2} + \mathcal{I}_{\mu,UZ}\right]\mathcal{I}_{\mu,UU}^{3/4}}\right)^{1/6}\label{B.mu.X.AI.Paper}\\
  h^{S}_{\mu,Z}&=\left(\frac{\mathcal{I}_{\mu,UU}}{\mathcal{I}_{\mu,ZZ}}\right)^{1/4}h^{S}_{\mu,U},\text{ where}\label{B.mu.Z.AI.Paper}
\end{align}
\begin{equation*}
  \begin{array}{l}
   Q_{\mu,1}=\int\left(\gamma(u,u,z)+\sigma^2_\epsilon\right)\,d(u,z),\;\;
   \mathcal{I}_{\mu,UZ}=\int\mu^{(2,0)}(u,z)\mu^{(0,2)}(u,z)\,f_{UZ}(u,z)\,d(u,z),\\
   \mathcal{I}_{\mu,UU}=\int\left(\mu^{(2,0)}(u,z)\right)^2\,f_{UZ}(u,z)\,d(u,z),\,\text{and}\;\;
   \mathcal{I}_{\mu,ZZ}=\int\left(\mu^{(0,2)}(u,z)\right)^2\,f_{UZ}(u,z)\,d(u,z).
  \end{array}
\end{equation*}
\end{theorem}

\begin{theorem}[Sparse - optimal bandwidths for $\hat{\gamma}$]\label{AMISE-I_opt_BDW_gamma}
Let $\sqrt{M}/n^{1/5}\to 0$ and $(u_1,u_2,z)$ be an interior point of $[0,1]^3$. Under our setup the AMISE optimal bandwidths for the LLK estimator in Eq.~\eqref{Estimator_gamma} are then given by
\begin{align}
h^{S}_{\gamma,U}&=\left(\frac{R(K_\gamma)\;Q_{\gamma,1}\;4\;\sqrt{2}\;\mathcal{I}_{\gamma,ZZ}^{3/2}}{nM\,\left(\nu_{2}(K_\gamma)\right)^2\,\left(2\,\left(\nu_{2}(K_\gamma)\right)^2\,\mathcal{I}_{\gamma,U_{(1)}Z} + C_{\mathcal{I}}\right)\,\left(C_{\mathcal{I}}-\mathcal{I}_{\gamma,U_{(1)}Z}\right)^{3/2}}\right)^{1/7}\label{B.gamma.X.AI.Paper}\\
h^{S}_{\gamma,Z}&=\left(\frac{C_{\mathcal{I}}-\mathcal{I}_{\gamma,U_{(1)}Z}}{2\,\mathcal{I}_{\gamma,ZZ}}\right)^{1/2}h^{S}_{\gamma,U},\label{B.gamma.Z.AI.Paper}
\end{align}
\begin{equation*}
\begin{array}{rcl}
\text{where \quad}C_{\mathcal{I}}&=&(\mathcal{I}_{\gamma,U_{(1)}Z}^{2} + 4 \,
  (\mathcal{I}_{\gamma,U_{(1)}U_{(1)}}+\mathcal{I}_{\gamma,U_{(1)}U_{(2)}})\,
  \mathcal{I}_{\gamma,ZZ})^{1/2},\\
Q_{\gamma,1}&=&\int\left(\tilde{\gamma}((u_1,u_2),(u_1,u_2),z)+\sigma^2_\varepsilon(u_1,u_2,z)\right)\,d(u_1,u_2,z)\\
    \mathcal{I}_{\gamma,U_{(1)}U_{(1)}}&=&\int\left(\gamma^{(2,0,0)}(u_1,u_2,z)\right)^2\,f_{UUZ}(u_1,u_2,z)\,d(u_1,u_2,z),\\
    \mathcal{I}_{\gamma,U_{(1)}U_{(2)}}&=&\int\left(\gamma^{(2,0,0)}(u_1,u_2,z)\gamma^{(0,2,0)}(u_1,u_2,z)\right)\,f_{UUZ}(u_1,u_2,z)\,d(u_1,u_2,z),\\
    \mathcal{I}_{\gamma,U_{(1)}Z}
    &=&\int\gamma^{(2,0,0)}(u_1,u_2,z)\gamma^{(0,0,2)}(u_1,u_2,z)\,f_{UUZ}(u_1,u_2,z)\,d(u_1,u_2,z),
    \quad\text{and}\\
    \mathcal{I}_{\gamma,ZZ}   &=&\int\left(\gamma^{(0,0,2)}(u_1,u_2,z)\right)^2\,f_{UUZ}(u_1,u_2,z)\,d(u_1,u_2,z).
\end{array}
\end{equation*}
\end{theorem}

The bandwidth rates are well-known for bi- and trivariate nonparametric estimators and essentially equivalent results can be found, e.g., in \cite{herrmann1995bandwidth}. The superscript S stands for sparse covariate adjusted functional data.

The following Corollaries \ref{C_AN_mu} and \ref{C_AN_gamma} contain our asymptotic normality results for the estimators $\hat{\mu}$ and $\hat{\gamma}$ for sparse functional data.

\begin{corollary}[Sparse - asymptotic normality of $\hat{\mu}$]\label{C_AN_mu}
Let $m/n^{1/5}\to 0$, let $(u,z)$ be an interior point of $[0,1]^2$, and assume optimal bandwidth choices. Under our setup the LLK estimator in Eq.~\eqref{Estimator_mu} is then asymptotically normal.\\
\textit{(a)} Without finite sample correction:
\begin{align*}
\left(\frac{\hat{\mu}(u,z;h_{\mu,U}^{S},h_{\mu,Z}^{S})-B_\mu(u,z;h_{\mu,U}^{S},h_{\mu,Z}^{S})-\mu(u,z)}{\sqrt{V_\mu^{I}(u,z;h_{\mu,U}^{S},h_{\mu,Z}^{S})}}\right)\overset{a}{\sim}N(0,1)
\end{align*}
\textit{(b)} With finite sample correction:
\begin{align*}
\left(\frac{\hat{\mu}(u,z;h_{\mu,U}^{S},h_{\mu,Z}^{S})-B_\mu(u,z;h_{\mu,U}^{S},h_{\mu,Z}^{S})-\mu(u,z)}{\sqrt{V_\mu^{I}(u,z;h_{\mu,U}^{S},h_{\mu,Z}^{S})+V_\mu^{II}(u,z;h_{\mu,Z}^{S})}}\right)\overset{a}{\sim}N(0,1)
\end{align*}
\end{corollary}

\begin{corollary}[Sparse - asymptotic normality of $\hat{\gamma}$]\label{C_AN_gamma}
Let $\sqrt{M}/n^{1/5}\to 0$, let $(u_1,u_2,z)$ be an interior point of $[0,1]^3$, and assume optimal bandwidth choices. Under our setup the LLK estimator in Eq.~\eqref{Estimator_gamma} is then asymptotically normal.\\
\textit{(a)} Without finite sample correction:
\begin{align*}
\left(\frac{\hat{\gamma}(u_1,u_2,z;h_{\gamma,U}^{S},h_{\gamma,Z}^{S})-B_\gamma(u_1,u_2,z;h_{\gamma,U}^{S},h_{\gamma,Z}^{S})-\gamma(u_1,u_2,z)}{\sqrt{V_\gamma^{I}(u_1,u_2,z;h_{\gamma,U}^{S},h_{\gamma,Z}^{S})}}\right)
\overset{a}{\sim}N(0,1)
\end{align*}
\textit{(b)} With finite sample correction:
\begin{align*}
\left(\frac{\hat{\gamma}(u_1,u_2,z;h_{\gamma,U}^{S},h_{\gamma,Z}^{S})-B_\gamma(u_1,u_2,z;h_{\gamma,U}^{S},h_{\gamma,Z}^{S})-\gamma(u_1,u_2,z)}{\sqrt{V_\gamma^{I}(u_1,u_2,z;h_{\gamma,U}^{S},h_{\gamma,Z}^{S})+V_\gamma^{II}(u_1,u_2,z;h_{\gamma,Z}^{S})}}\right)
\overset{a}{\sim}N(0,1)
\end{align*}
\end{corollary}

The above corollaries imply that the standard optimal convergence rates for bivariate ($nm^{-1/3}$) and trivariate ($nM^{-2/7}$) LLK estimators are attained. Corollaries \ref{C_AN_mu} \textit{(a)} and \ref{C_AN_gamma} \textit{(a)} are essentially equivalent to Theorems 3.2 and 3.4 of \cite{jiang2010covariate} who, however, consider the LLK estimators under the finite-$m$ asymptotic. In contrast, we show that these results hold for all $n\to\infty$ and $m,M\to\infty$ with $m/n^{1/5}\to 0$ and $\sqrt{M}/n^{1/5}\to 0$ respectively. Corollaries \ref{C_AN_mu} \textit{(b)} and \ref{C_AN_gamma} \textit{(b)} contain our finite-sample corrections that allow for robust inferences; see our simulation study in Section \ref{sec:Sim}.

\subsection{Dense functional data}\label{SecondVarTerm}
If the second variance summands, $V_\mu^{II}$ and $V_\gamma^{II}$, are the leading variance terms, it is possible to achieve \emph{uni}variate convergence rates for the bi- and trivariate estimators $\hat{\mu}(u,z)$ and $\hat{\gamma}(u_1,u_2,z)$. By contrast to the preceding section, however, it is impossible to determine the optimal bandwidths by using only the leading variance terms $V_\mu^{II}$ and $V_\gamma^{II}$ respectively. The trick is to determine the bandwidth expressions in a hierarchical manner: The optimal $Z$-bandwidths $h_{\mu,Z}^{D}$ and $h_{\gamma,Z}^{D}$ must be derived by optimizing with respect to the leading (i.e., $Z$-related) bias and variance terms. Given the optimal $Z$-bandwidths, the optimal $U$-bandwidths $h_{\mu,U}^{D}$ and $h_{\gamma,U}^{D}$ can be determined by optimizing the subsequent lower-order bias and variance terms. This leads to the following optimal bandwidth expressions, where the superscript D suggests that we are considering the case of dense covariate adjusted functional data.

\begin{theorem}[Dense - optimal bandwidths for $\hat{\mu}$]\label{AMISE-II_opt_BDW_mu}
Let $m/n^{1/5}\to\infty$ and $(u,z)$ be an interior point of $[0,1]^2$. Under our setup the AMISE optimal bandwidths for the LLK estimator in Eq.~\eqref{Estimator_mu} are then given by
\begin{align}
h^{D}_{\mu,Z}&=\left(\frac{R(\kappa)\,Q_{\mu,2}}{n\,\left(\nu_{2}(K_\mu)\right)^2\,\mathcal{I}_{\mu,ZZ}}\right)^{1/5}\quad\text{and}\label{h.mu.Z.AMISE2}\\
h^{D}_{\mu,U}&=\left(\frac{R(K_\mu)\,Q_{\mu,1}}{nm\,\left(\nu_{2}(K_\mu)\right)^2\,\mathcal{I}_{\mu,UZ}}\right)^{1/3}\left(h^{D}_{\mu,Z}\right)^{-1},\quad\text{where}\label{h.mu.U.AMISE2}
\end{align}
\begin{equation*}
\begin{array}{l}
Q_{\mu,1}=\int\left(\gamma(u,u,z)+\sigma^2_\epsilon\right)\,d(u,z),\;
\mathcal{I}_{\mu,UZ}=\int \mu^{(2,0)}(u,z)\mu^{(0,2)}(u,z)\,f_{UZ}(u,z)\,d(u,z),\\
Q_{\mu,2}=\int\gamma(u,u,z)\,f_U(u)\,d(u,z),\,\text{and}\;\;
\mathcal{I}_{\mu,ZZ}=\int(\mu^{(0,2)}(u,z))^2\,f_{UZ}(u,z)\,d(u,z).
\end{array}
\end{equation*}
\end{theorem}

\begin{theorem}[Dense - optimal bandwidths for $\hat{\gamma}$]\label{AMISE-II_opt_BDW_gamma}
Let $\sqrt{M}/n^{1/5}\to\infty$ and $(u_1,u_2,z)$ be an interior point of $[0,1]^3$. Under our setup the AMISE optimal bandwidths for the LLK estimator in Eq.~\eqref{Estimator_gamma} are then given by
\begin{align}
h^{D}_{\gamma,Z}&=\left(\frac{R(\kappa)\,Q_{\gamma,2}}{n\,\left(\nu_{2}(K_\gamma)\right)^2\,\mathcal{I}_{\gamma,ZZ}}\right)^{1/5}\label{h.gamma.Z.AMISE2}\quad\text{and}\\
h^{D}_{\gamma,U}&=\left(\frac{R(K_\gamma)\,Q_{\gamma,1}}{nM\,\left(\nu_{2}(K_\gamma)\right)^2\,\mathcal{I}_{\gamma,U_{(1)}Z}}\right)^{1/4}\left(h^{D}_{\gamma,Z}\right)^{-3/4},\label{h.gamma.U.AMISE2}
\end{align}
\begin{equation*}
\begin{array}{rcl}
\text{where}\quad Q_{\gamma,1}&=&\int\left(\tilde{\gamma}((u_1,u_2),(u_1,u_2),z)+\sigma^2_\varepsilon(u_1,u_2,z)\right)\,d(u_1,u_2,z),\\
Q_{\gamma,2}&=&\int\tilde{\gamma}((u_1,u_2),(u_1,u_2),z)\,f_{UU}(u_1,u_2)\,d(u_1,u_2,z),\\
\mathcal{I}_{\gamma,ZZ}&=&\int(\gamma^{(0,0,2)}(u_1,u_2,z))^2\,f_{UUZ}(u_1,u_2,z)\,d(u_1,u_2,z),\quad\text{and}\\
\mathcal{I}_{\gamma,U_{(1)}Z}&=&\int\gamma^{(2,0,0)}(u_1,u_2,z)\gamma^{(0,0,2)}(u_1,u_2,z)\,f_{UUZ}(u_1,u_2,z)\,d(u_1,u_2,z).
\end{array}
\end{equation*}
\end{theorem}

Note that the $\AMISE$ optimal bandwidths $h_{\mu,U}^D$ and $h_{\mu,Z}^D$ in Eqs.~\eqref{h.mu.Z.AMISE2} and \eqref{h.mu.U.AMISE2} and $h_{\gamma,U}^D$ and $h_{\gamma,Z}^D$ in Eqs.~\eqref{h.gamma.Z.AMISE2} and \eqref{h.gamma.U.AMISE2} are in a sense \emph{anti}-proportional to each other. A larger $Z$-bandwidth implies a smaller $U$-bandwidth, and vice versa, for given $n$ and $m$. This is contrary to the classical multiple bandwidth results where the single bandwidths are directly proportional to each other.

To explain this finding, observe that a larger $Z$-bandwidth implies that more functions $X_i(.,Z_i)$ are used for computing local averages. However, taking averages over an increased amount of data reduces variance so that we can afford some further increase in variance by using \emph{under}smoothing bandwidths in $U$-direction. This undersmoothing strategy leads to a better estimation performance. A related result can be found in \cite{Benko2009a}, who, however, consider the simpler context without covariate adjustments.

The following corollaries contain our asymptotic normality result for the estimators $\hat{\mu}$ and $\hat{\gamma}$ in the case of dense functional data:
\begin{corollary}[Dense - asymptotic normality of $\hat{\mu}$]\label{C_AN_mu_2}
Let $m/n^{1/5}\to\infty$, let $(u,z)$ be an interior point of $[0,1]^2$, and assume optimal bandwidth choices. Under our setup the LLK estimator in Eq.~\eqref{Estimator_mu} is then asymptotically normal.\\
\textit{(a)} Without finite sample correction:
\begin{align*}
\left(\frac{\hat{\mu}(u,z;h_{\mu,U}^{D},h_{\mu,Z}^{D})-B^D_\mu(u,z;h_{\mu,Z}^{D})-\mu(u,z)}{\sqrt{V_\mu^{II}(u,z;h_{\mu,Z}^{D})}}\right)\overset{a}{\sim}N(0,1)
\end{align*}
\textit{(b)} With finite sample correction:
\begin{align*}
\left(\frac{\hat{\mu}(u,z;h_{\mu,U}^{D},h_{\mu,Z}^{D})-B^D_\mu(u,z;h_{\mu,Z}^{D})-\mu(u,z)}{\sqrt{V_\mu^{I}(u,z;h_{\mu,U}^{D},h_{\mu,Z}^{D})+V_\mu^{II}(u,z;h_{\mu,Z}^{D})}}\right)\overset{a}{\sim}N(0,1)
\end{align*}
where $B^D_\mu(u,z;h_{\mu,Z}^{D})=\frac{1}{2}\nu_2(K_\mu)(h_{\mu,Z}^{D})^2\mu^{(0,2)}(u,z)$.
\end{corollary}

\begin{corollary}[Dense - asymptotic normality of $\hat{\gamma}$]\label{C_AN_gamma_2}
Let $\sqrt{M}/n^{1/5}\to\infty$, let $(u_1,u_2,z)$ be an interior point of $[0,1]^3$, and assume optimal bandwidth choices. Under our setup the LLK estimator in Eq.~\eqref{Estimator_gamma} is then asymptotically normal.\\
\textit{(a)} Without finite sample correction:
\begin{align*}
\left(\frac{\hat{\gamma}(u_1,u_2,z;h_{\gamma,U}^{D},h_{\gamma,Z}^{D})-B^D_\gamma(u_1,u_2,z;h_{\gamma,Z}^{D})-\gamma(u_1,u_2,z)}{\sqrt{V_\gamma^{II}(u_1,u_2,z;h_{\gamma,Z}^{D})}}\right)\overset{a}{\sim}N(0,1)
\end{align*}
\textit{(b)} With finite sample correction:
\begin{align*}
\left(\frac{\hat{\gamma}(u_1,u_2,z;h_{\gamma,U}^{D},h_{\gamma,Z}^{D})-B^D_\gamma(u_1,u_2,z;h_{\gamma,Z}^{D})-\gamma(u_1,u_2,z)}{\sqrt{V_\gamma^{I}(u_1,u_2,z;h_{\gamma,U}^{D}+h_{\gamma,Z}^{D})+V_\gamma^{II}(u_1,u_2,z;h_{\gamma,Z}^{D})}}\right)\overset{a}{\sim}N(0,1),
\end{align*}
where $B^D_\gamma(u_1,u_2,z;h_{\gamma,Z}^{D})=\frac{1}{2}\nu_2(K_\gamma)(h_{\mu,Z}^{D})^2\gamma^{(0,0,2)}(u_1,u_2,z)$.
\end{corollary}

The above corollaries imply that the optimal convergence rate $(n^{-2/5})$ of \emph{uni}variate LLK estimators is attained, although, we are considering bi- and trivariate estimators $\hat\mu(u,z)$ and $\hat\gamma(u_1,u_2,z)$. Indeed, if $m/n^{1/5}\to \infty$ and $\sqrt{M}/n^{1/5}\to \infty$, the LLK estimators $\hat\mu(u,z)$ and $\hat\gamma(u_1,u_2,z)$ behave like LLK estimators for univariate regression functions with asymptotically negligible $U$-related bias and variance components and with $Z_i$ as their only covariate. That is, LLK estimators behave as if the sample of $n$ functions $\{X_1(.,Z_1),\dots,X_n(.,Z_n)\}$ were \emph{fully} observed such that smoothing needs to be done only in $Z$-direction.

This is qualitatively equivalent to the results in Corollaries 3.2 and 3.5 of \cite{zhang2016sparse}, who, however, consider the simpler context without covariate adjustments. Their LLK estimators behave as if they were classical parametric moment estimators applied to a sample of $n$ fully observed random functions without covariate adjustments.

\section{Simulation}\label{sec:Sim}
In order to assess the finite-sample properties of our asymptotic normality results, we consider the performance of the following pointwise confidence intervals:
\begin{align*}
\text{Sparse}&\text{ - without finite-sample correction (Corollary \ref{C_AN_mu} \textit{(a)}):}\\
&\operatorname{CI}^S(u,z)=\hat\mu^S_{\text{bc}}(u,z;h_{\mu,U}^{S},h_{\mu,Z}^{S})\pm
z_{1-\alpha/2}\sqrt{V_\mu^{I}(u,z;h_{\gamma,U}^{S},h_{\gamma,Z}^{S})}\\
\text{Sparse}&\text{ - with finite-sample correction (Corollary \ref{C_AN_mu} \textit{(b)}):}\\
&\operatorname{CI}^S_C(u,z)=\hat\mu^S_{\text{bc}}(u,z;h_{\mu,U}^{S},h_{\mu,Z}^{S})\pm
z_{1-\alpha/2}\sqrt{V_\mu^{I}(u,z;h_{\gamma,U}^{S},h_{\gamma,Z}^{S})+V_\mu^{II}(u,z;h_{\gamma,Z}^{S})}\\
\text{Dense}&\text{ - without finite-sample correction (Corollary \ref{C_AN_mu_2} \textit{(a)}):}\\
&\operatorname{CI}^D(u,z)=\hat\mu^D_{\text{bc}}(u,z;h_{\mu,U}^{D},h_{\mu,Z}^{D})\pm
z_{1-\alpha/2}\sqrt{V_\mu^{I}(u,z;h_{\gamma,U}^{D},h_{\gamma,Z}^{D})}\\
\text{Dense}&\text{ - with finite-sample correction (Corollary \ref{C_AN_mu_2} \textit{(b)}):}\\
&\operatorname{CI}^D_C(u,z)=\hat\mu^D_{\text{bc}}(u,z;h_{\mu,U}^{D},h_{\mu,Z}^{D})\pm
z_{1-\alpha/2}\sqrt{V_\mu^{I}(u,z;h_{\gamma,U}^{D},h_{\gamma,Z}^{D})+V_\mu^{II}(u,z;h_{\gamma,Z}^{D})},
\end{align*}
where $z_{1-\alpha/2}$ is the $(1-\alpha/2)$-quantile of the standard normal distribution and
\begin{equation}\label{BCor}
\begin{array}{rcl}
\hat \mu^S_{\text{bc}}(u,z;h^S_{\mu,U},h^S_{\mu,Z})&=&\hat{\mu}(u,z;h^S_{\mu,U},h^S_{\mu,Z})-B_\mu(u,z;h^S_{\mu,U},h^S_{\mu,Z})\\
\hat \mu^D_{\text{bc}}(u,z;h^D_{\mu,U},h^D_{\mu,Z})&=&\hat{\mu}(u,z;h^D_{\mu,U},h^D_{\mu,Z})-B^D_\mu(u,z;h^D_{\mu,Z})
\end{array}
\end{equation}
denote the bias-corrected mean estimates.

The above theoretical confidence intervals are infeasible as they depend on the unknown bandwidth, bias, and variance expressions. To achieve feasible confidence intervals we replace the unknown theoretical bandwidth parameters ($h^S_{\mu,U}$, $h^S_{\mu,Z}$, $h^D_{\mu,U}$, and $h^D_{\mu,Z}$) using simple but effective rule-of-thumb estimates ($\hat h^S_{\mu,U}$, $\hat h^S_{\mu,Z}$, $\hat h^D_{\mu,U}$, and $\hat h^D_{\mu,Z}$), based on our theoretical bandwidth expressions as described in Section \ref{sec:Bdw_approx}. The unknown bias ($B_\mu$ and $B^D_\mu$) and variance ($V_\mu^{I}$ and $V_\mu^{II}$) terms are estimated using LLK estimators ($\hat B_\mu$, $\hat B^D_\mu$, $\hat V_\mu^{I}$, and $\hat V_\mu^{II}$) as described in Section \ref{sec:bias_var}.

We simulate data from $Y_{ij}=\mu(U_{ij},Z_i)+X^c_i(U_{ij},Z_i)+\epsilon_{ij}$, where
$U_{ij}\sim\text{Unif}(0,1)$,
$Z_i\sim \text{Unif}(0,1)$,
$\epsilon_{ij}\sim\text{N}(0,0.05^2)$,
$X_i^c(u,z)=\xi_{i1}\psi_1(u,z)+\xi_{i2}\psi_2(u,z)$,
$\xi_{i1}\sim\text{N}(0,\lambda_1)$,
$\xi_{i2}\sim\text{N}(0,\lambda_2)$,
$\gamma(u_1,u_2,z)=\lambda_1\psi_1(u_1,z)\psi_1(u_2,z)+\lambda_2\psi_2(u_1,z)\psi_2(u_2,z)$, $m\in\{5,10,15\}$, and $n=100$. The following two Data Generating Processes (DGPs) are used, where DGP 2 is essentially that of \cite{jiang2010covariate}:

\bigskip
\begin{tabular}{lll}
\textbf{DGP 1}& Meanfunction:  & $\mu(u,z)=5\sin(\pi u z/2)$\\
              & Eigenfunctions:& $\psi_1(u,z)=\sin(\pi u z)$\\
              &                & $\psi_2(u,z)=\sin(2\pi u z)$\\
              & Eigenvalues:   & $\lambda_1=3$\text{ and }$\lambda_2=2$\\\\
\textbf{DGP 2}& Meanfunction:  & $\mu(u,z)=u-u\sin(\pi z)+z \cos(\pi u)$\\
              & Eigenfunctions:& $\psi_1(u,z)=-\cos(\pi (u+z/2))\sqrt{2}$\\ &                & $\psi_2(u,z)= \sin(\pi(u+z/2))\sqrt{2}$\\
              & Eigenvalues:   & $\lambda_1=z/9$\text{ and }$\lambda_2=z/36$\\
\end{tabular}
\bigskip

For each DGP and each sample size combination we draw 5000 Monte-Carlo repetitions and compute the empirical coverage probabilities of the pointwise confidence intervals at the following three $(u,z)$-points:
\begin{center}
$\text{Point 1}=(0.25,0.75)
 \text{,\; Point 2}=(0.5, 0.5)
 \text{,\; and \;Point 3}=(0.75,0.25)$
\end{center}

We focus on confidence intervals for the mean function; nonparametric confidence intervals for the covariance function are typically not used in practice as they involve the nonparametric estimation of the fourth-moment function $\tilde\gamma$ contained in the unknown variance terms $V_\gamma^{I}$ and $V_\gamma^{II}$. The latter is complicated and typically leads to very unstable estimates due to an accumulation of preceding estimation errors. Our theoretical results on the LLK estimator $\hat\gamma$ are, nevertheless, of crucial importance for estimating the unknown variance expressions $V$ of the confidence intervals (see Section \ref{sec:bias_var}). For evaluating the estimation results with respect to the covariance function, we consider the average integrated squared error. The simulation study was conducted using a standard PC and lasted about five days.
\begin{figure}[!htb]
    \centering
    \includegraphics[width=1\textwidth]{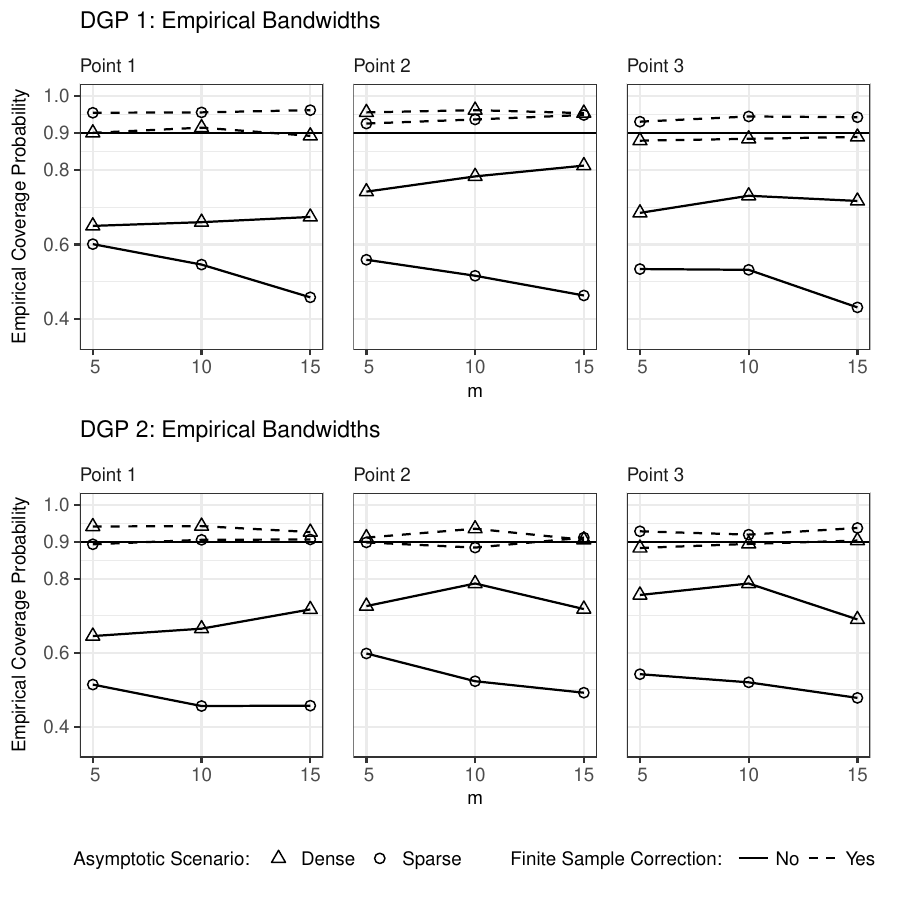}
  \caption[]{Empirical coverage probabilities of the feasible and infeasible confidence intervals.}
  \label{fig:Sim}
\end{figure}

\begin{figure}[!htb]
    \centering
    \includegraphics[width=1\textwidth]{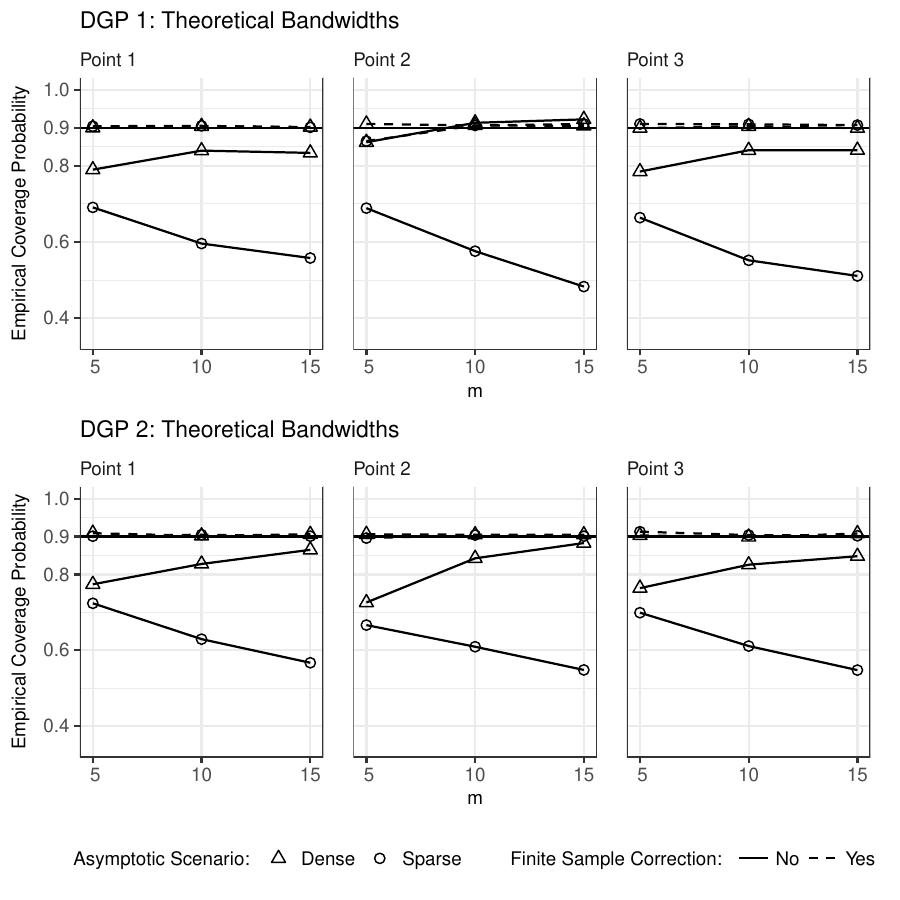}
  \caption[]{Empirical coverage probabilities of the feasible and infeasible confidence intervals.}
  \label{fig:Sim_Th}
\end{figure}

The two top panels in Figure \ref{fig:Sim} show the empirical coverage probabilities of the feasible confidence intervals with plugged-in estimates $\hat h^S_{\mu,U}$, $\hat h^S_{\mu,Z}$, $\hat h^D_{\mu,U}$, $\hat h^D_{\mu,Z}$, $\hat B_\mu$, $\hat B^D_\mu$, $\hat V^{I}_\mu$, and $\hat V^{II}_\mu$ from Sections \ref{sec:Bdw_approx} and \ref{sec:bias_var}. The two bottom panels show the empirical coverage probabilities of the infeasible theoretical confidence intervals based on the theoretical bandwidth, bias, and variance expressions. The infeasible confidence intervals serve as validating benchmarks, since they allow us to abstract from the additional estimation errors that are due to the plug-in estimates. We use $(1-\alpha/2)=0.9$ as our nominal coverage probability.

Let us first consider the feasible version of the sparse confidence interval $\operatorname{CI}^S$. This is an interesting special case, since essentially the same confidence interval would be used by a practitioner who takes the asymptotic normality result in Theorem 3.2 of \cite{jiang2010covariate} as a theoretical basis. This confidence interval shows a very poor performance with far to small coverage probabilities. The problem occurs in our sparse data scenario with sample sizes $m=5$ and $n=100$ and -- as expected -- becomes worse as $m$ increases.

By contrast, the feasible version of the dense confidence interval $\operatorname{CI}^D$ performs quite satisfactorily. Though, the best and most stable results are achieved by the feasible versions of the confidence intervals with finite-sample corrections, $\operatorname{CI}^S_C$ and $\operatorname{CI}^D_C$, both showing an almost equally good performance. All of our results on the feasible confidence intervals are essentially equivalent to those for the infeasible theoretical benchmark confidence intervals shown in the two bottom panels in Figure \ref{fig:Sim_Th}. This comparison serves as a validation of our simulation results, since it shows that the results are not driven by bad and too imprecise plug-in estimates.

\begin{table}[!ht]
\centering
\caption{Empirical versus theoretical variances. The figures in the upper panel are based on the sparse bandwidths, $\hat h^S_{\mu,U}$ and $\hat h^S_{\mu,Z}$, from Eq.s \eqref{B.mu.X.AI.ROT.Paper} and \eqref{B.mu.Z.AI.ROT.Paper}; those of the lower panel are based on the dense bandwidths, $\hat h^D_{\mu,U}$ and $\hat h^D_{\mu,Z}$, from Eq.s \eqref{h.mu.Z.ROT.AMISE2} and \eqref{h.mu.U.ROT.AMISE2}.}
\label{tab:1}
\begin{tabular}{r@{/}l c ccc c ccc}
\toprule
\multicolumn{2}{l}{} &&\multicolumn{3}{c}{DGP-1} && \multicolumn{3}{c}{DGP-2}\\\cline{4-6}\cline{8-10}
\multicolumn{2}{l}{Point 1:\;$(u,z)=(0.25,0.75)$}&&m=5&m=10&m=15 &&m=5&m=10&m=15\\\hline
\multicolumn{9}{l}{}\\[-1.9ex]
$\widehat{\text{Var}}\big(\hat\mu(u,z)\big)$&$V^{I}_\mu(u,z)$
&&2.9&3.3&6.7  &&3.1&3.5&4.1\\
$\widehat{\text{Var}}\big(\hat\mu(u,z)\big)$&$\big(V^{I}_\mu(u,z)+V^{II}_\mu(u,z)\big)$
&&1.2&1.3&1.5 &&1.0&0.9&0.9\\[1.1ex]
\midrule
\multicolumn{9}{l}{}\\[-1.9ex]
$\widehat{\text{Var}}\big(\hat\mu(u,z)\big)$&$V^{II}_\mu(u,z)$
&&1.7&1.3&1.6  &&1.8&1.3&1.5\\
$\widehat{\text{Var}}\big(\hat\mu(u,z)\big)$&$\big(V^{I}_\mu(u,z)+V^{II}_\mu(u,z)\big)$
&&0.9&1.1&1.1  &&1.0&0.9&1.0\\
\bottomrule
\end{tabular}
\end{table}

The reason for the poor performance of the confidence interval $\operatorname{CI}^S$ is shown in the first row of Table \ref{tab:1}. The variance term $V_\mu^{I}=V^{I}_\mu(u,z)$ used to construct $\operatorname{CI}^S$, severely underestimates the finite-sample variance $\widehat{\text{Var}}(\hat\mu)=\widehat{\text{Var}}(\hat\mu(u,z))$ of the LLK estimator $\hat\mu(u,z)$, where $\widehat{\text{Var}}(\hat\mu)$ is computed from the 5000 Monte Carlo replications. For $m=5$, the first variance term, $V^{I}_\mu$, is $2.9$ and $3.1$ times smaller than the actual finite-sample variance $\widehat{\text{Var}}(\hat\mu)$ and -- as expected -- the ratio becomes worse as $m$ increases. This leads to too narrow confidence intervals $\operatorname{CI}^S$ and hence to too small coverage probabilities. This observation is in line with the relatively small $\theta=1/5$ threshold for differentiating between sparse and dense functional data. The small $\theta=1/5$ threshold implies that the functional-data-specific second variance term $V_\mu^{II}$ will be non-negligible in real data scenarios where $m$ is relatively small in comparison to $n^{1/5}$. Therefore, including the second variance terms $V^{II}_\mu(u,z)$ leads to strongly improved approximations of the finite-sample variances of $\hat\mu$; see the second and fourth row in Table \ref{tab:1}. This explains the superior performances of the confidence intervals, $\operatorname{CI}^S_C$ and $\operatorname{CI}^D_C$, incorporating our finite-sample corrections. 

For evaluating the estimation results with respect to the covariance function, we consider the average integrated squared error $\int_0^1\int_0^1\int_0^1\big(\hat{\gamma}(u_1,u_2,z)-\gamma(u_1,u_2,z)\big)^2du_1du_2dz$ based on 100 Monte Carlo simulations using the sparse and dense rule-of-thumb bandwidth approximations of Section \ref{sec:ROT}. Table \ref{tab:2} shows that the sparse and dense bandwidths perform both comparably well. The performance of the sparse bandwidths gets slightly worse as $m$ increases and the performance of the dense bandwidths slightly improve as $m$ increases.
\begin{table}[!ht]
\centering
\caption{Average integrated squared error $\int_0^1\int_0^1\int_0^1\big(\hat{\gamma}(u_1,u_2,z)-\gamma(u_1,u_2,z)\big)^2du_1du_2dz$.}
\label{tab:2}
\begin{tabular}{llccc}
\toprule
    & Bandwidths &  $m=5$ & $m=10$ & $m=15$ \\ 
\midrule
DGP1& Sparse     & 0.0012 & 0.0015 & 0.0020 \\ 
    & Dense      & 0.0015 & 0.0012 & 0.0007 \\ \hline
DGP2& Sparse     & 0.0015 & 0.0016 & 0.0022 \\
    & Dense      & 0.0019 & 0.0017 & 0.0012 \\ 
\bottomrule
\end{tabular}
\end{table} 

\section{Bandwidth, bias and variance approximations}\label{sec:ROT}
\subsection{Rule-of-thumb bandwidth approximations}\label{sec:Bdw_approx}
Our above bandwidth expressions are infeasible as they depend on the unknown quantities
$\mathcal{I}_{\mu,UU}$,
$\mathcal{I}_{\mu,UZ}$, $\mathcal{I}_{\mu,ZZ}$, $Q_{\mu,1}$, $Q_{\mu,2}$,
$\mathcal{I}_{\gamma,U_{(1)}U_{(1)}}$, $\mathcal{I}_{\gamma,U_{(1)}Z}$,
$\mathcal{I}_{\gamma,ZZ}$, $Q_{\gamma,1}$, and $Q_{\gamma,2}$.
Following \cite{Fan1996}, we suggest approximating them using global polynomial regression models. In the following we list our rule-of-thumb approximations for the bandwidths in Eq.s \eqref{B.mu.X.AI.Paper}-\eqref{h.gamma.U.AMISE2}:\\
\noindent Sparse rule-of-thumb bandwidths for $\hat\mu$:
\begin{align}
\hat{h}^{S}_{\mu,U}&=\left(\frac{R(K_\mu)\,Q_{\mu,1}\,\hat{\mathcal{I}}_{\mu_{\text{poly}},ZZ}^{3/4}}{nm\,\left(\nu_{2}(K_\mu)\right)^2\,\left[\hat{\mathcal{I}}_{\mu_{\text{poly}},UU}^{1/2}\,\hat{\mathcal{I}}_{\mu,ZZ}^{1/2} + \hat{\mathcal{I}}_{\mu_{\text{poly}},UZ}\right]\hat{\mathcal{I}}_{\mu_{\text{poly}},UU}^{3/4}}\right)^{1/6}\label{B.mu.X.AI.ROT.Paper}\\
\hat{h}^{S}_{\mu,Z}&=\left(\frac{\hat{\mathcal{I}}_{\mu_{\text{poly}},UU}}{\hat{\mathcal{I}}_{\mu_{\text{poly}},ZZ}}\right)^{1/4}\hat{h}^{S}_{\mu,U}\label{B.mu.Z.AI.ROT.Paper}
\end{align}
\noindent Sparse rule-of-thumb bandwidths for $\hat\gamma$:
\begin{align}
\hat{h}^{S}_{\gamma,U}&=\left(\frac{R(K_\gamma)\;\hat{Q}_{\gamma_{\text{poly}},1}\;4\;\sqrt{2}\;\hat{\mathcal{I}}_{\gamma_{\text{poly}},ZZ}^{3/2}}{nM\,\left(\nu_{2}(K_\gamma)\right)^2\,\left(2\,\left(\nu_{2}(K_\gamma)\right)^2\,\hat{\mathcal{I}}_{\gamma_{\text{poly}},U_{(1)}Z} + \hat{C}_{\mathcal{I}}\right)\,\left(\hat{C}_{\mathcal{I}}-\hat{\mathcal{I}}_{\gamma_{\text{poly}},U_{(1)}Z}\right)^{3/2}}\right)^{1/7}\label{B.gamma.X.AI.ROT.Paper}\\
\hat{h}^{S}_{\gamma,Z}&=\left(\frac{\hat{C}_{\mathcal{I}}-\hat{\mathcal{I}}_{\gamma_{\text{poly}},U_{(1)}Z}}{2\,\hat{\mathcal{I}}_{\gamma,ZZ}}\right)^{1/2}\hat{h}^{S}_{\gamma,U},\;\text{where}\label{B.gamma.Z.AI.ROT.Paper}\\
\hat{C}_{\mathcal{I}}&=(\hat{\mathcal{I}}_{\gamma_{\text{poly}},U_{(1)}Z}^{2} + 4 \,(\hat{\mathcal{I}}_{\gamma_{\text{poly}},U_{(1)}U_{(1)}}+\hat{\mathcal{I}}_{\gamma_{\text{poly}},U_{(1)}U_{(2)}})\,
  \hat{\mathcal{I}}_{\gamma_{\text{poly}},ZZ})^{1/2}.\notag
\end{align}
\noindent Dense rule-of-thumb bandwidths for $\hat\mu$:
\begin{align}
\hat{h}^{D}_{\mu,Z}&=\left(\frac{R(\kappa)\,\hat{Q}_{\mu_{\text{poly}},2}}{n\,\left(\nu_{2}(K_\mu)\right)^2\,\hat{\mathcal{I}}_{\mu_{\text{poly}},ZZ}}\right)^{1/5}\label{h.mu.Z.ROT.AMISE2}\\
\hat{h}^{D}_{\mu,U}&=\left(\frac{R(K_\mu)\,\hat{Q}_{\mu_{\text{poly}},1}}{nm\,\left(\nu_{2}(K_\mu)\right)^2\,\hat{\mathcal{I}}_{\mu_{\text{poly}},UZ}}\right)^{1/3}\left(\hat{h}^{D}_{\mu,Z}\right)^{-1}\label{h.mu.U.ROT.AMISE2}
\end{align}
\noindent Dense rule-of-thumb bandwidths for $\hat\gamma$:
\begin{align}
\hat{h}^{D}_{\gamma,Z}&=\left(\frac{R(\kappa)\,\hat{Q}_{\gamma_{\text{poly}},2}}{n\,\left(\nu_{2}(K_\gamma)\right)^2\,\hat{\mathcal{I}}_{\gamma_{\text{poly}},ZZ}}\right)^{1/5}\label{h.gamma.Z.ROT.AMISE2}\\
\hat{h}^{D}_{\gamma,U}&=\left(\frac{R(K_\gamma)\,\hat{Q}_{\gamma_{\text{poly}},1}}{nM\,\left(\nu_{2}(K_\gamma)\right)^2\,\hat{\mathcal{I}}_{\gamma_{\text{poly}},U_{(1)}Z}}\right)^{1/4}\left(\hat{h}^{D}_{\gamma_{\text{poly}},Z}\right)^{-3/4}\label{h.gamma.U.ROT.AMISE2}
\end{align}

The above rule-of-thumb bandwidth expressions are based on the following estimates for the sparse rule-of-thumb bandwidths:
\begin{equation*}
\begin{array}{rcl}
\hat{\mathcal{I}}_{\mu_\text{poly},UU}&=&\int_{\supp(f_{UZ})}(\hat{\mu}_{\text{poly}}^{(2,0)}(u,z))^2\,\hat{f}_{UZ}(u,z)\,d(u,z),\\[1ex]
\hat{\mathcal{I}}_{\mu_\text{poly},UZ}&=&\int_{\supp(f_{UZ})}\hat{\mu}_{\text{poly}}^{(2,0)}(u,z)\hat{\mu}_{\text{poly}}^{(0,2)}(u,z)\,\hat{f}_{UZ}(u,z)\,d(u,z),\\[1ex]
\hat{\mathcal{I}}_{\mu_\text{poly},ZZ}&=&\int_{\supp(f_{UZ})}(\hat{\mu}_{\text{poly}}^{(0,2)}(u,z))^2\,\hat{f}_{UZ}(u,z)\,d(u,z),\\[1ex]
\hat{Q}_{\mu_\text{poly},1}&=&\int_{\supp(f_{UZ})}\hat{\gamma}^{\text{ND}}_{\text{poly}}(u,u,z)\,d(u,z),\\[1ex]
\hat{Q}_{\mu_\text{poly},2}&=&\int_{\supp(f_{UZ})}\hat{\gamma}_\text{poly}(u,u,z)\,\hat{f}_U(u)\,d(u,z),
\end{array}
\end{equation*}
and for the dense rule-of-thumb bandwidths:
\begin{equation*}
\begin{array}{rcl}
\hat{\mathcal{I}}_{\gamma_{\text{poly}},U_{(1)} U_{(1)}}&=&\int_{\supp(f_{UZ})}(\hat{\gamma}_{\text{poly}}^{(2,0,0)}(u,u,z))^2\,\hat{f}_{UZ}(u,z)\,d(u,z),\\[1ex]
\hat{\mathcal{I}}_{\gamma_\text{poly},U_{(1)}Z}&=&\int_{\supp(f_{UUZ})}\hat{\gamma}_{\text{poly}}^{(2,0,0)}(u,u,z)\hat{\gamma}_{\text{poly}}^{(0,0,2)}(u,u,z)\,\hat{f}_{UUZ}(u,u,z)\,d(u,u,z),\\[1ex]
\hat{\mathcal{I}}_{\gamma_\text{poly},ZZ}&=&\int_{\supp(f_{UUZ})}(\hat{\gamma}_{\text{poly}}^{(0,0,2)}(u,u,z))^2\,\hat{f}_{UUZ}(u,u,z)\,d(u,u,z),\\[1ex]
\hat{Q}_{\gamma_\text{poly},1}&=&\int_{\supp(f_{UUZ})}\hat{\tilde{\gamma}}^{\text{ND}}_\text{poly}((u_1,u_2),(u_1,u_2),z)\,d(u_1,u_2,z),\quad\text{and}\\[1ex]
\hat{Q}_{\gamma_\text{poly},2}&=&\int_{\supp(f_{UUZ})}\hat{\tilde{\gamma}}_\text{poly}((u_1,u_2),(u_1,u_2),z)\;\hat{f}_{UU}(u_1,u_2)\,d(u_1,u_2,z).
\end{array}
\end{equation*}
The estimates $\hat{\mu}_{\text{poly}}$, $\hat{\gamma}^{\text{ND}}_{\text{poly}}$,
$\hat{\gamma}_{\text{poly}}$,
$\hat{\tilde{\gamma}}^{\text{ND}}_{\text{poly}}$,
$\hat{\tilde{\gamma}}_{\text{poly}}$, $\hat{\mu}_{\text{poly}}^{(2,0)}$,
$\hat{\mu}_{\text{poly}}^{(0,2)}$, $\hat{\gamma}_{\text{poly}}^{(2,0,0)}$, and $\hat{\gamma}_{\text{poly}}^{(0,0,2)}$ are the ordinary least squares estimates (and their derivatives) of the following polynomial regression models:

\bigskip
\noindent$\boldsymbol{\mu}_{\textbf{poly}}\textbf{:}$ The model
$\mu_{\text{poly}}(u,z)$ is fitted via regressing $Y_{ij}$ on powers (each
up to the fourth power) of
$U_{ij}$, $Z_i$, and $U_{ij}\cdot Z_i$ for all $i\in\{1,\dots,n\}$ and $j\in\{1,\dots,m\}$, i.e.,
$Y_{ij}=\mu_{\text{poly}}(U_{ij},Z_i)+\text{error}_{ij}$,
where
$\mu_{\text{poly}}(U_{ij},Z_i)=\beta_0 +
\sum_{q=1}^4\left(
\beta^{U}_q  U_{ij}^q +
\beta^{Z}_q  Z_{ij}^q +
\beta^{UZ}_q (U_{ij}Z_i)^q
\right).$

\bigskip
\noindent$\boldsymbol{\gamma}_{\textbf{poly}}$\textbf{:} The model
$\gamma_{\text{poly}}(u_1,u_2,z)$ is fitted via regressing $C^{\text{poly}}_{ijk}$ on powers (each up to the fourth power) of $U_{ij}$, $U_{ik}$, $Z_i$, $U_{ij}\cdot Z_i$, and
$U_{ik}\cdot Z_i$ for all $i\in\{1,\dots,n\}$ and all $j,k\in\{1,\dots,m\}$ with $j\neq k$, i.e.,
$C^{\text{poly}}_{ijk}=\gamma_{\text{poly}}(U_{ij},U_{ik},Z_i)+\text{error}_{ijk}$,
where
$C^{\text{poly}}_{ijk}=(Y_{ij}-\mu_{\text{poly}}(U_{ij},Z_i))(Y_{jt}-\mu_{\text{poly}}(U_{ik},Z_i))$ and \\
$\gamma_{\text{poly}}(U_{ij},U_{ik},Z_i)=\beta_0 +
\sum_{q=1}^4\big(
\beta^{U,1}_q U_{ij}^q +
\beta^{U,2}_q U_{ik}^q +
\beta^{Z}_q   Z_i^q    +
\beta^{UZ,1}_q (U_{ij}Z_i)^q +
\beta^{UZ,2}_q (U_{ik}Z_i)^q
\big).$

\bigskip
\noindent$\boldsymbol{\tilde{\gamma}}_{\textbf{poly}}$\textbf{:} The model
$\tilde\gamma_{\text{poly}}((u_1,u_2),(u_3,u_4),z)$ is fitted via regressing
$\mathbb{C}^{\text{poly}}_{ijk\ell m}$
on powers (each
up to the fourth power) of $U_{ij}$, $U_{ik}$, $U_{i\ell}$, $U_{im}$, and $Z_i$ for all
$i\in\{1,\dots,n\}$ and all $j,k,\ell,m\in\{1,\dots,m\}$ such that ($j\neq\ell$ AND $k\neq m$), i.e.,
$\mathbb{C}^{\text{poly}}_{ijk\ell m}=\tilde\gamma_{\text{poly}}((U_{ij},U_{ik}),(U_{i\ell},U_{im}),Z_i)+\text{error}_{ijk\ell m}$, where
$\mathbb{C}^{\text{poly}}_{ijk\ell m}=(C^{\text{poly}}_{ijk}-\gamma_{\text{poly}}(U_{ij},U_{ik},Z_i))(C^{\text{poly}}_{i\ell m}-\gamma_{\text{poly}}(U_{i\ell},U_{im},Z_i))$ and \\
$\tilde\gamma_{\text{poly}}((U_{ij},U_{ik}),(U_{i\ell},U_{im}),Z_i)
=\beta_0 +
\sum_{q=1}^4\left(
\beta_q^{U,1} U_{ij}^q +
\beta_q^{U,2} U_{ik}^q +
\beta_q^{U,3} U_{i\ell}^q +
\beta_q^{U,4} U_{im}^q    +
\beta_q^{Z} Z_i^q         +\right.$
$\left.+\beta_q^{UZ,1} (U_{ij}Z_i)^q +
\beta_q^{UZ,2} (U_{ik}Z_i)^q +
\beta_q^{UZ,3} (U_{i\ell}Z_i)^q +
\beta_q^{UZ,4} (U_{im}Z_i)^q
\right).$

\bigskip
\noindent$\boldsymbol{\gamma}^{\textbf{S}}_{\textbf{poly}}$\textbf{:} The model $\gamma^{\text{ND}}_{\text{poly}}(u,u,z)$
is fitted via
regressing the noise-contaminated
diagonal values $C^{\text{poly}}_{ij}$ on powers (each
up to the fourth power)
of $U_{ij}$, and $Z_i$ for all $i\in\{1,\dots,n\}$ and $j\in\{1,\dots,m\}$, i.e.,
$C^{\text{poly}}_{ij}=\gamma^{\text{ND}}_{\text{poly}}(U_{ij},U_{ij},Z_i)+\text{error}_{ij}$, where
$C^{\text{poly}}_{ij}=(Y_{ij}-\mu_{\text{poly}}(U_{ij},Z_i))^2$ and
$\gamma^{\text{ND}}_{\text{poly}}(U_{ij},U_{ij},Z_i)
=\beta_0 +
\sum_{q=1}^4\left(
\beta^U_q U_{ij}^q +
\beta^Z_q    Z_i^q    +
\beta^{UZ}_q (U_{ij}Z_i)^q
\right).$ The ND in $\gamma^{\text{ND}}_{\text{poly}}$ suggest that we are estimating the noise-contaminated diagonal values $\gamma(u,u,z)+\sigma_\epsilon$.

\bigskip
\noindent$\boldsymbol{\tilde{\gamma}}^{\textbf{S}}_{\textbf{poly}}$\textbf{:} The model
$\tilde{\gamma}^{\text{ND}}_{\text{poly}}(u_1,u_2,z)$
is fitted via regressing the noise-contaminated diagonal values
$\mathbb{C}_{ijk}^{\text{poly}}$ on powers (each
up to the fourth power) of
$U_{ij}$, $U_{ik}$, and $Z_i$ for all $i\in\{1,\dots,n\}$, and $j,k\in\{1,\dots,m\}$, i.e.,
$\mathbb{C}^{\text{poly}}_{ijk}=\tilde{\gamma}^{\text{ND}}_{\text{poly}}(U_{ij},U_{ik},Z_i)+\text{error}_{ijk}$, where
$\mathbb{C}^{\text{poly}}_{ijk}=(C^{\text{poly}}_{ijk}-\gamma_{\text{poly}}(U_{ij},U_{ik},Z_i))^2$ and
$\tilde{\gamma}^{\text{ND}}_{\text{poly}}(U_{ij},U_{ik},Z_i)=\beta_0 +
\sum_{q=1}^4\left(
\beta_q^{U,1} U_{ij}^q +
\beta_q^{U,2} U_{ik}^q +
\beta_q^{Z} Z_i^q
\right).$ The ND in $\tilde{\gamma}^{\text{ND}}_{\text{poly}}$ suggest that we are estimating the noise-contaminated diagonal values $\tilde{\gamma}((u_1,u_2),(u_1,u_2),z)+\sigma_\varepsilon(u_1,u_2,z)$.

Estimates of the densities $f_{UZ}$ and $f_{UUZ}$ are computed as kernel density estimates using Gaussian kernels and bandwidth determined by cross-validation.

\paragraph{Remark} It is important to specify the models $\boldsymbol{\mu_{\text{poly}}}$ and $\boldsymbol{\gamma_{\text{poly}}}$ using interaction terms, since otherwise their partial derivatives $\hat{\mu}_{\text{poly}}^{(2,0)}$, $\hat{\mu}_{\text{poly}}^{(0,2)}$, $\hat{\gamma}_{\text{poly}}^{(2,0,0)}$, and $\hat{\gamma}_{\text{poly}}^{(0,0,2)}$ would degenerate.

\subsection{Bias and variance estimates}\label{sec:bias_var}
Following \cite{hardle1988bootstrapping}, we approximate the unknown second derivatives $\mu^{(2,0)}$ and $\mu^{(0,2)}$ in $B_\mu$ and $B^D_\mu$ using local polynomial estimators. That is, we approximate $B_\mu(u,z;h_{\mu,U},h_{\mu,Z})$ and $B^D_\mu(u,z;h_{\mu,Z})$ by
{\begin{equation*}
\begin{array}{rcl}
\hat{B}_\mu(u,z;h_{\mu,U},h_{\mu,Z})&=&\frac{\nu_2(K_\mu)}{2}
\Big(h_{\mu,U}^2\hat{\mu}^{(2,0)}\big(u,z;g_{\mu,U},g_{\mu,Z}\big)+
h_{\mu,Z}^2\hat{\mu}^{(0,2)}\big(u,z;g_{\mu,U},g_{\mu,Z}\big)\Big)\\[2ex]
\text{and}\quad\hat{B}^D_\mu(u,z;h_{\mu,Z})&=&\frac{\nu_2(K_\mu)}{2}
h_{\mu,Z}^2\hat{\mu}^{(0,2)}\big(u,z;g_{\mu,U},g_{\mu,Z}\big)
\end{array}
\end{equation*}}
where $\hat{\mu}^{(2,0)}$ and $\hat{\mu}^{(0,2)}$ are local polynomial (order $3$) kernel estimators of $\mu^{(2,0)}$ and $\mu^{(0,2)}$:
\begin{equation*}
\begin{array}{l}
\hat{\mu}^{(2,0)}(u,z;g_{\mu,U},g_{\mu,Z})=
2!\,e_3^{\top}\left([\mathbf{1},\mathbf{U}_u^{1:3},\mathbf{Z}_z^{1:3}]^{\top}\mathbf{W}_{\mu,uz}[\mathbf{1},\mathbf{U}_u^{1:3},\mathbf{Z}_z^{1:3}]\right)^{-1}[\mathbf{1},\mathbf{U}_u^{1:3},\mathbf{Z}_z^{1:3}]^{\top}\mathbf{W}_{\mu,uz}\mathbf{Y}\\
\hat{\mu}^{(0,2)}(u,z;g_{\mu,U},g_{\mu,Z})=
2!\,e_6^{\top}\left([\mathbf{1},\mathbf{U}_u^{1:3},\mathbf{Z}_z^{1:3}]^{\top}\mathbf{W}_{\mu,uz}[\mathbf{1},\mathbf{U}_u^{1:3},\mathbf{Z}_z^{1:3}]\right)^{-1}[\mathbf{1},\mathbf{U}_u^{1:3},\mathbf{Z}_z^{1:3}]^{\top}\mathbf{W}_{\mu,uz}\mathbf{Y},
\end{array}
\end{equation*}
with $e_3^{\top}=(0,0,1,0,0,0,0)$, $e_6^{\top}=(0,0,0,0,0,1,0)$, $\mathbf{U}_u^{1:3}=[\mathbf{U}_u,\mathbf{U}_u^2,\mathbf{U}_u^3]$,  $\mathbf{Z}_z^{1:3}=[\mathbf{Z}_z,\mathbf{Z}_z^2,\mathbf{Z}_z^3]$, and
$\mathbf{W}_{\mu,uz}=\operatorname{diag}(\dots,g^{-1}_{\mu,U}\,\kappa(g^{-1}_{\mu,U}(U_{ij}-u))\;
g^{-1}_{\mu,Z}\,\kappa(g^{-1}_{\mu,Z}(Z_i-z)),\dots)$.

For estimating the bandwidths $g_{\mu,U}$ and $g_{\mu,Z}$ we use bivariate GCV based on second-order differences. We follow the procedure of \cite{CS15}, but use a GCV-penalty instead of their proposed (asymptotically equivalent) $C_p$-penalty.

For approximating the unknown noise-contaminated diagonal of the covariance function, $\gamma(u,u,z)+\sigma_\epsilon^2$, contained in $V_\mu^{I}$, we propose to use a LLK estimator. That is, we approximate $V_\mu^{I}$ by
\begin{align*}
\hat{V}_\mu^{I}(u,z;h_{\mu,U},h_{\mu,Z},h_{\gamma,U},h_{\gamma,Z})&=(nm)^{-1}\left[(h_{\mu,U}h_{\mu,Z})^{-1}R(K_\mu)\frac{\hat{\gamma}^{\text{ND}}(u,u,z;h_{\gamma,U},h_{\gamma,Z})}{\hat{f}_{UZ}(u,z)}\,\right]
\end{align*}
where the estimator of the Noisy Diagonal (ND), $\hat{\gamma}^{\text{ND}}(u,u,z;h_{\gamma,U},h_{\gamma,Z})\approx \{\gamma(u,u,z)+\sigma_\epsilon^2\}$, is defined as the following LLK estimator:
\begin{align*}
\hat{\gamma}^{\text{ND}}(u,u,z;h_{\gamma,U},h_{\gamma,Z})=e_1^{\top}\left([\mathbf{1},\mathbf{U}_u,\mathbf{Z}_z]^{\top}\mathbf{W}_{\gamma,uz}[\mathbf{1},\mathbf{U}_u,\mathbf{Z}_z]\right)^{-1}[\mathbf{1},\mathbf{U}_{u},\mathbf{Z}_z]^{\top}\mathbf{W}_{\gamma,uz}\hat{\mathbf{C}},
\end{align*}
where $\hat{\mathbf{C}}=(\hat{C}_{111},\dots,C_{ijj}\dots,\hat{C}_{nmm})^{\top}$ consists only of the diagonal raw-covariances, i.e., $\hat{C}_{ijj}=(Y_{ij}-\hat{\mu}(U_{ij},Z_i;h_{\mu,U},h_{\mu,Z}))^2$. Note that $\hat{\gamma}^{\text{ND}}$ is equivalent to the LLK estimator $\hat{V}$ in \cite{jiang2010covariate}.

Finally, for estimating the unknown quantity $\gamma(u,u,z)$ in $V_\mu^{II}$ we can use our LLK estimator $\hat{\gamma}(u,u,z;h_{\gamma,U},h_{\gamma,Z})$ as defined in \eqref{Estimator_gamma}. That is we estimate $V_\mu^{II}$ by
\begin{align*}
\hat{V}_\mu^{II}(u,z;h_{\mu,Z},h_{\gamma,U},h_{\gamma,Z})&=n^{-1}\left[\left(\frac{m-1}{m}\right)(h_{\mu,Z})^{-1}\,R(\kappa)\frac{\hat{\gamma}(u,u,z;h_{\gamma,U},h_{\gamma,Z})}{\hat{f}_Z(z)}\right].
\end{align*}

The bandwidths for the LLK estimators $\hat\gamma$ and $\hat\gamma^{\text{ND}}$ are selected according to our Rule-of-thumb bandwidth approximations in Eq.s \eqref{B.gamma.X.AI.ROT.Paper}, \eqref{B.gamma.Z.AI.ROT.Paper}, \eqref{h.gamma.Z.ROT.AMISE2}, and \eqref{h.gamma.U.ROT.AMISE2}. An alternative approach to the above proposed rule-of-thumb approximations might be to completely adapt the bootstrap procedure of \cite{hardle1988bootstrapping}, which is, however, computationally more demanding.

\section{Application}\label{sec:App}
\begin{figure}[!hbt]
\centering
\includegraphics[width=1\textwidth]{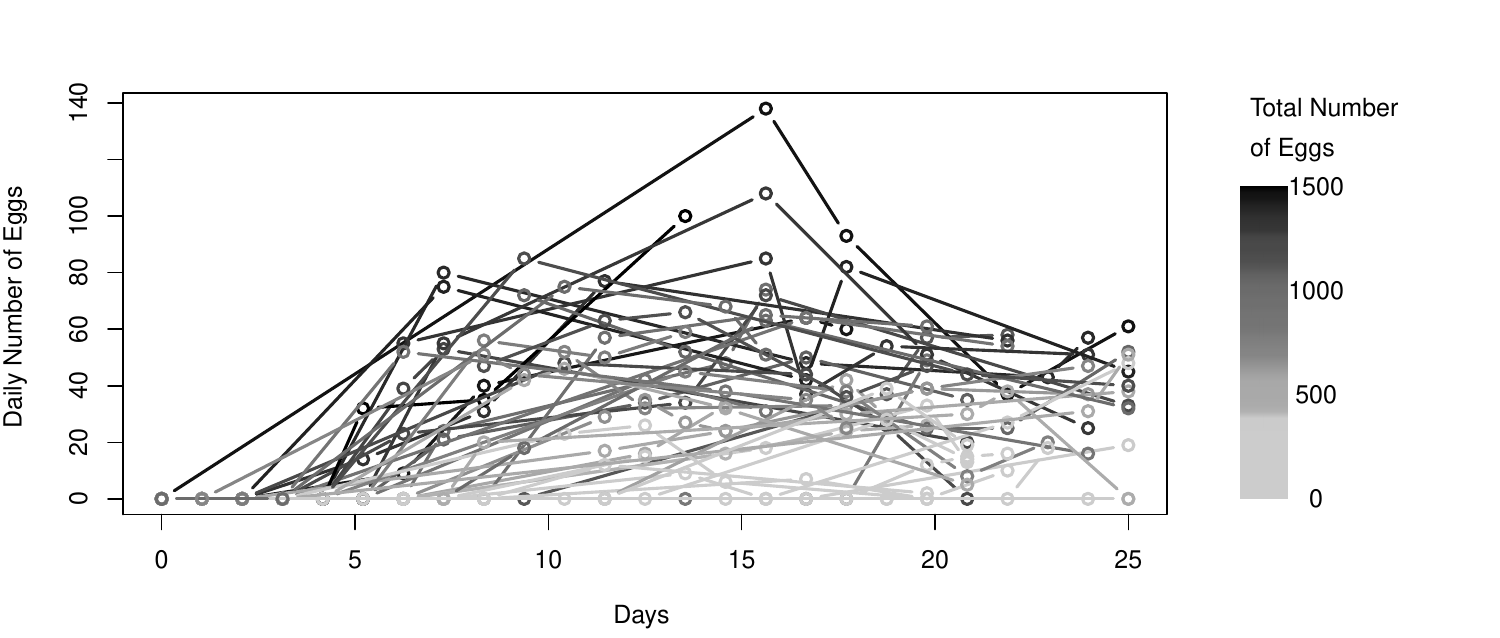}
\caption[]{Mediterranean fruit flies data originally analyzed by \cite{CLM98}.}
\label{fig:Dat} 
\end{figure}

\begin{figure}[!htb]
\centering
\includegraphics[width=1\textwidth]{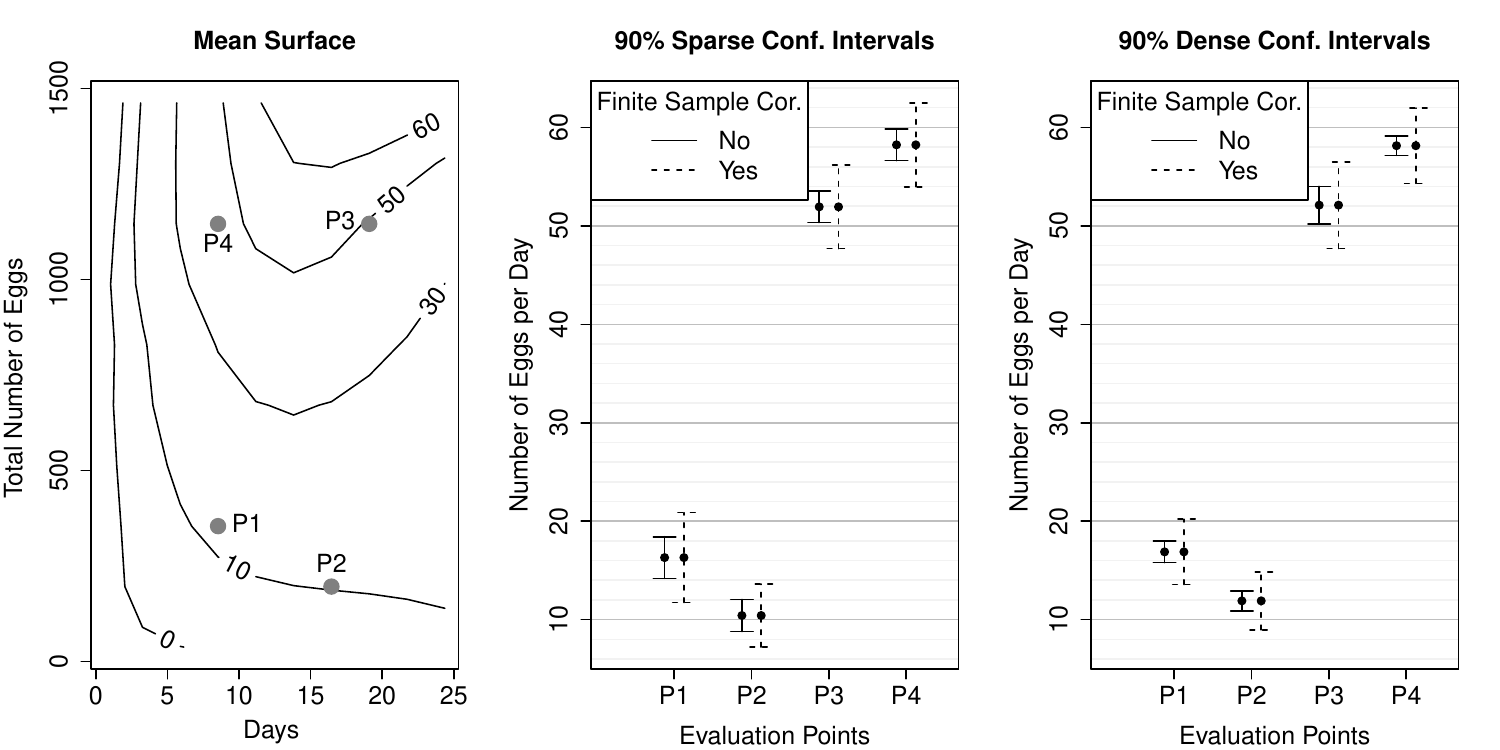}
\caption[]{{\sc Left Panel:} Contour plot of the estimated mean function. {\sc Middle and Right Panel:} 90\% sparse and dense confidence intervals without finite sample correction (solid lines) and with finite sample correction (dashed line).}
\label{fig:Res}
\end{figure}

We consider the well-known reproductive data for Mediterranean fruit flies (Ceratitis capitata) as provided in the R-package fdapace of \cite{FDAPACE}. This is a subsample of the data previously analyzed in \cite{CLM98} containing the daily numbers of eggs laid from 789 medflies during the first 25 days of their lives. As \cite{jiang2010covariate}, we construct a sparse data set by randomly selecting without replacement $m=10$ observations from the 25 measurements of each fruit fly. Random selections of $m=5$ or $m=15$ observations lead to qualitatively equivalent results. Following the original analysis in \cite{CLM98}, we analyze the relationship between the daily reproduction $Y_{ij}$, measured at day $U_{ij}$, and the total reproduction $Z_i$. Figure \ref{fig:Dat} shows the data, where we only display a subset of 50 randomly selected trajectories to prevent an overcrowded and unclear plot. The left panel in Figure \ref{fig:Res} shows the contour plot of the surface of the estimated mean function using our sparse bandwidths and essentially replicates Figure 7 of the original analysis in \cite{CLM98}. The contour lines clearly indicate a dependency between the daily mean reproduction and the total mean reproduction of medflies. 

In order to showcase the relevance of our theoretical results, we select four points, \textsf{P1}, \textsf{P2}, \textsf{P3}, and \textsf{P4}, and compare their confidence intervals based on the sparse ($\operatorname{CI}^S_C$ and $\operatorname{CI}^S$) and dense ($\operatorname{CI}^D_C$ and $\operatorname{CI}^D$) asymptotic scenarios with and without finite sample correction (see middle and right panel of Figure \ref{fig:Res}). A Bonferroni correction is used to adjust for the multiple testing. From our simulation study we know that the variance components of the confidence intervals without finite sample corrections tend to underestimate the actual variance of the nonparametric mean estimator. That is, inference based on the confidence intervals without finite sample corrections will have a tendency for over-rejection of the null-hypotheses of equal means, due to an under-estimated pointwise variance component in finite samples. 

This adverse effect can be seen when using the confidence intervals in order to check for significant differences between \textsf{P1} vs.~\textsf{P2} and \textsf{P3} vs.~\textsf{P4}. The confidence intervals without finite sample corrections ($\operatorname{CI}^S$ and $\operatorname{CI}^D$) are extremely narrow and suggest significant differences between the means at \textsf{P1} vs.~\textsf{P2} and \textsf{P3} vs.~\textsf{P4}. These significant differences are quite implausible, since the points \textsf{P1} and \textsf{P2} as well as \textsf{P3} and \textsf{P4} lie almost at the same contour lines. By contrast, the confidence intervals with our finite sample corrections ($\operatorname{CI}_C^S$ and $\operatorname{CI}_C^D$) are considerably wider and do not indicate such implausibly significant differences. Qualitatively similar results can be showcased, e.g., for $m=5$ and $m=15$; however, their display is omitted in order not to unnecessarily prolong the manuscript. 

\section*{Acknowledgements}
I want to thank Alois Kneip (University of Bonn) and Piotr Kokoszka (Colorado State University) for fruitful discussions and valuable comments which helped to improve this research work. Additionally, I want to thank Ir\`ene Gijbels (KU Leuven) and Jane-Ling Wang (UC Davis) for valuable comments on my talk at the CMStatistics in 2015 where I presented an earlier version of this manuscript. Many thanks go to the student assistants of the Institute of Financial Economics and Statistics of the University of Bonn for coding assistance.
 
\appendix
 
\section{Proofs}
\subsection{Proof of Theorem \ref{Bias_and_Variance_mu}}
Proof of Theorem \ref{Bias_and_Variance_mu}, part \textit{(i)}: For simplicity, consider a second-order kernel function $\kappa$ with compact support such as the Epanechnikov kernel. This is, of course, without loss of generality, but allows for a more compact proof. Define $\mathbf{H}_\mu=\mydiag(h^2_{\mu,U},h^2_{\mu,Z})$, $\mathbf{U}=(U_{11},\dots,U_{nm})^\top$, and $\mathbf{Z}=(Z_1,\dots,Z_n)^\top$. Using a Taylor-expansion of $\mu$ around $(u,z)$, the conditional bias of the estimator $\hat{\mu}(u,z;\mathbf{H})$ can be written as
\begin{align}
\E(\hat{\mu}(u,z;\mathbf{H}_\mu)-\mu(u,z)|\mathbf{U},\mathbf{Z})=\frac{1}{2}e_1^{\top}\left((nm)^{-1}[\mathbf{1},\mathbf{U}_{u},\mathbf{Z}_{z}]^{\top}\mathbf{W}_{\mu,uz}[\mathbf{1},\mathbf{U}_{u},\mathbf{Z}_{z}]\right)^{-1}\times\label{Bias1}\\
\times(nm)^{-1}[\mathbf{1},\mathbf{U}_{u},\mathbf{Z}_{z}]^{\top}\mathbf{W}_{\mu,uz}\left(\boldsymbol{\mathcal{Q}}_\mu(u,z)+\mathbf{R}_\mu(u,z)\right),\notag
\end{align}
where $\boldsymbol{\mathcal{Q}}_\mu(u,z)$ is a $nm\times 1$ vector with typical elements
\begin{eqnarray*}
  (U_{ij}-u,Z_i-z)\boldsymbol{\mathcal{H}}_\mu(u,z)(U_{ij}-u,Z_i-z)^{\top}\in\mathbb{R}
\end{eqnarray*}
with $\boldsymbol{\mathcal{H}}_\mu(u,z)$ being the Hessian matrix of the regression function $\mu(u,z)$. The $nm\times 1$ vector $\mathbf{R}_\mu(u,z)$ holds the remainder terms as in \cite{ruppert1994}.

Next we derive asymptotic approximations for the $3\times 3$ matrix\\
$\left((nm)^{-1}[\mathbf{1},\mathbf{U}_{u},\mathbf{Z}_{z}]^{\top} \mathbf{W}_{\mu,uz}[\mathbf{1},\mathbf{U}_{u},\mathbf{Z}_{z}]\right)^{-1}$
and the $3\times 1$ matrix $(nm)^{-1}[\mathbf{1},\mathbf{U}_{u},\mathbf{Z}_{z}]^{\top}
\mathbf{W}_{\mu,uz}\boldsymbol{\mathcal{Q}}_\mu(u,z)$ of the right hand side of
Eq.~\eqref{Bias1}. Using standard procedures from kernel density
estimation it is easy to derive that \\
$(nm)^{-1}[\mathbf{1},\mathbf{U}_{u},\mathbf{Z}_{z}]^{\top}\mathbf{W}_{\mu,uz}[\mathbf{1},\mathbf{U}_{u},\mathbf{Z}_{z}]=$
\begin{eqnarray*}
  \left(\begin{matrix}
      f_{UZ}(u,z)+o_p(1)                                      &\quad \nu_{2}(K_\mu)\mathbf{D}_{f_{UZ}}(u,z)^{\top}\mathbf{H}_\mu+o_p(\mathbf{1}^\top\mathbf{H}_\mu)\\
      \nu_{2}(K_\mu)\mathbf{H}_\mu \mathbf{D}_{f_{UZ}}(u,z)+o_p(\mathbf{H}_\mu\mathbf{1})&\quad \nu_{2}(K_\mu)\mathbf{H}_\mu f_{UZ}(u,z)+o_p(\mathbf{H}_\mu)\\
    \end{matrix}\right),
\end{eqnarray*}
where $\mathbf{1}=(1,1)^\top$ and $\mathbf{D}_{f_{UZ}}(u,z)$ is the vector of first order partial derivatives (i.e., the gradient) of the pdf $f_{UZ}$ at $(u,z)$. Inversion of the above block matrix yields
\begin{eqnarray}\label{pre_results}
  \left((nm)^{-1}[\mathbf{1},\mathbf{U}_{u},\mathbf{Z}_{z}]^{\top}\mathbf{W}_{\mu,uz}[\mathbf{1},\mathbf{U}_{u},\mathbf{Z}_{z}]\right)^{-1}=&&
\end{eqnarray}
{\small\begin{eqnarray*}
  \left(\begin{matrix}
      \left(f_{UZ}(u,z)\right)^{-1}+o_p(1)                          &\quad  -\mathbf{D}_{f_{UZ}}(u,z)^{\top}\left(f_{UZ}(u,z)\right)^{-2}+o_p(\mathbf{1}^\top)\\
      -\mathbf{D}_{f_{UZ}}(u,z)\left(f_{UZ}(u,z)\right)^{-2}+o_p(\mathbf{1}) &\quad \left(\nu_{2}(K_\mu)\mathbf{H}_\mu f_{UZ}(u,z)\right)^{-1}+o_p(\mathbf{H}_\mu)\\
\end{matrix}\right).
\end{eqnarray*}}
The $3\times 1$ matrix $(nm)^{-1}[\mathbf{1},\mathbf{U}_{u},\mathbf{Z}_{z}]^{\top}\mathbf{W}_{\mu,uz}\boldsymbol{\mathcal{Q}}_\mu(u,z)$ can be partitioned as following:
\begin{eqnarray*}
  (nm)^{-1}[\mathbf{1},\mathbf{U}_{u},\mathbf{Z}_{z}]^{\top}\mathbf{W}_{\mu,uz}\boldsymbol{\mathcal{Q}}_\mu(u,z)&=&\left(\begin{matrix}\texttt{upper element}\\
      \texttt{lower bloc}
    \end{matrix}\right),
\end{eqnarray*}
where the $1\times 1$ dimensional \texttt{upper element} can be approximated by
{\small\begin{align}
  &(nm)^{-1}\sum_{ij}K_{\mu,h}(U_{ij}-u,Z_{i}-z)(U_{ij}-u,Z_{i}-z)\boldsymbol{\mathcal{H}}_\mu(u,z)(U_{ij}-u,Z_{i}-z)^{\top}\label{Q1}\\
  =&\left(\nu_{2}(\kappa)\right)^2tr\left\{\mathbf{H}_\mu\boldsymbol{\mathcal{H}}_\mu(u,z)\right\}f_{UZ}(u,z)+o_p(tr(\mathbf{H}_{\mu}))\notag
\end{align}}
and the $2\times 1$ dimensional \texttt{lower bloc} is equal to
\begin{align}
&(nm)^{-1}\sum_{ij}\left\{K_{\mu,h}(U_{ij}-u,Z_{i}-z)(U_{ij}-u,Z_{i}-z)\boldsymbol{\mathcal{H}}_\mu(u,z)(U_{ij}-u,Z_i-z)^{\top}\right\}\times\label{Q2}\\
&\times(U_{ij}-u,Z_{i}-z)^{\top}=O_p(\mathbf{H}_\mu^{3/2}\mathbf{1}).\notag
\end{align}\noindent
Plugging the approximations of Eqs.~\eqref{pre_results}-\eqref{Q2} into the first summand of the conditional bias expression in Eq.~\eqref{Bias1} leads to the following expression
\begin{align*}
&\frac{1}{2}e_1^{\top}((nm)^{-1}[\mathbf{1},\mathbf{U}_{u},\mathbf{Z}_{z}]^{\top}\mathbf{W}_{\mu,uz}[\mathbf{1},\mathbf{U}_{u},\mathbf{Z}_{z}])^{-1}(nm)^{-1}[\mathbf{1},\mathbf{U}_{u},\mathbf{Z}_{z}]^{\top}\mathbf{W}_{\mu,uz}\boldsymbol{\mathcal{Q}}_\mu(u,z)=\\
&=\frac{1}{2}\left(\nu_{2}(\kappa)\right)^2tr\left\{\mathbf{H}_\mu\boldsymbol{\mathcal{H}}_\mu(u,z)\right\}+o_p(tr(\mathbf{H}_{\mu})).
\end{align*}
Furthermore, it is easily seen that the second summand of the conditional bias expression in Eq.~\eqref{Bias1}, which holds the remainder term, is given by
\begin{align*}
\frac{1}{2}e_1^{\top}((nm)^{-1}[\mathbf{1},\mathbf{U}_{u},\mathbf{Z}_{z}]^{\top}\mathbf{W}_{\mu,uz}[\mathbf{1},\mathbf{U}_{u},\mathbf{Z}_{z}])^{-1}(nm)^{-1}[\mathbf{1},\mathbf{U}_{u},\mathbf{Z}_{z}]^{\top}\mathbf{W}_{\mu,uz}\mathbf{R}_\mu(u,z)=o_p(tr(\mathbf{H}_\mu)).
\end{align*}\noindent
Summation of the two latter expressions yields the asymptotic approximation of the conditional bias
\begin{align*}
  \E(\hat{\mu}(u,z;\mathbf{H}_\mu)-\mu(u,z)|\mathbf{U},\mathbf{Z})=\frac{1}{2}\left(\nu_{2}(\kappa)\right)^2tr\left\{\mathbf{H}_\mu \boldsymbol{\mathcal{H}}_\mu(u,z)\right\}+o_p(tr(\mathbf{H}_\mu)).
\end{align*}\noindent
This is our bias statement of Theorem \ref{Bias_and_Variance_mu} part
\textit{(i)}.

\bigskip

Proof of Theorem \ref{Bias_and_Variance_mu}, part \textit{(ii)}: In the following we derive the conditional variance of the local linear estimator $\V(\hat{\mu}(u,z;\mathbf{H}_\mu)|\mathbf{U},\mathbf{Z})=$
{\small\begin{align}
  =&e_1^{\top}([\mathbf{1},\mathbf{U}_{u},\mathbf{Z}_{z}]^{\top}\mathbf{W}_{\mu,uz}[\mathbf{1},\mathbf{U}_{u},\mathbf{Z}_{z}])^{-1}\,
  [\mathbf{1},\mathbf{U}_{u},\mathbf{Z}_{z}]^{\top}\mathbf{W}_{\mu,uz}\, \mathrm{Cov}(\mathbf{Y}|\mathbf{U},\mathbf{Z})\, \mathbf{W}_{\mu,uz}[\mathbf{1},\mathbf{U}_{u},\mathbf{Z}_{z}]\times \notag\\
  &\times ([\mathbf{1},\mathbf{U}_{u},\mathbf{Z}_{z}]^{\top}\mathbf{W}_{\mu,uz}[\mathbf{1},\mathbf{U}_{u},\mathbf{Z}_{z}])^{-1}u_1\notag\\\label{Varexpr}\\
  =&e_1^{\top}((nm)^{-1}[\mathbf{1},\mathbf{U}_{u},\mathbf{Z}_{z}]^{\top}\mathbf{W}_{\mu,uz}[\mathbf{1},\mathbf{U}_{u},\mathbf{Z}_{z}])^{-1}\times\,\notag\\
  &\times ((nm)^{-2}[\mathbf{1},\mathbf{U}_{u},\mathbf{Z}_{z}]^{\top}\mathbf{W}_{\mu,uz}\, \mathrm{Cov}(\mathbf{Y}|\mathbf{U},\mathbf{Z})\, \mathbf{W}_{\mu,uz}[\mathbf{1},\mathbf{U}_{u},\mathbf{Z}_{z}])\times \notag\\
  &\times ((nm)^{-1}[\mathbf{1},\mathbf{U}_{u},\mathbf{Z}_{z}]^{\top}\mathbf{W}_{\mu,uz}[\mathbf{1},\mathbf{U}_{u},\mathbf{Z}_{z}])^{-1}u_1,\notag 
\end{align}}
where $\mathrm{Cov}(\mathbf{Y}|\mathbf{U},\mathbf{Z})$ is the $nm\times nm$ matrix with typical elements
\begin{align*}
  &\mathrm{Cov}(Y_{ij},Y_{\ell k}|U_{ij},U_{\ell k},Z_i,Z_\ell)=\gamma_{|i-\ell|}((U_{ij},Z_i),(U_{\ell k},Z_\ell))+\sigma_\epsilon^2\mathbbm{1}{\left(i=\ell \text{ and }j=k\right)}
\end{align*}
with $\mathbbm{1}(.)$ being the indicator function.

We begin with analyzing the $3\times 3$ matrix
\begin{eqnarray*}
  (nm)^{-2}[\mathbf{1},\mathbf{U}_{u},\mathbf{Z}_{z}]^{\top}\mathbf{W}_{\mu,uz}\, \mathrm{Cov}(\mathbf{Y}|\mathbf{U},\mathbf{Z})\, \mathbf{W}_{\mu,uz}[\mathbf{1},\mathbf{U}_{u},\mathbf{Z}_{z}]
\end{eqnarray*}
using the following three Lemmas \ref{upper_left}-\ref{lower_right}.

\begin{lemma}\label{upper_left}
The upper-left scalar (block) of the matrix \\
$(nm)^{-2}[\mathbf{1},\mathbf{U}_{u},\mathbf{Z}_{z}]^{\top}\mathbf{W}_{\mu,uz}\mathrm{Cov}(\mathbf{Y}|\mathbf{U},\mathbf{Z})\mathbf{W}_{\mu,uz}[\mathbf{1},\mathbf{U}_{u},\mathbf{Z}_{z}]$ is given by
{\small\begin{eqnarray*}
  &&(nm)^{-2}\mathbf{1}^{\top}\mathbf{W}_{\mu,uz}\mathrm{Cov}(\mathbf{Y}|\mathbf{U},\mathbf{Z})\mathbf{W}_{\mu,uz}\mathbf{1}\\
  &=&(nm)^{-1}f_{UZ}(u,z)|\mathbf{H}_\mu|^{-1/2}R(K_\mu)\left(\gamma(u,u,z)+\sigma^2_\epsilon\right)(1+O_p(tr(\mathbf{H}_\mu^{1/2})))\\
  &+&n^{-1}(f_{UZ}(u,z))^2\left[\left(\frac{m-1}{m}\right)h_{\mu,Z}^{-1}R(\kappa)\frac{\gamma(u,u,z)}{f_Z(z)}\right]
  (1+O_p(tr(H^{1/2})))\\
  &=&O_p((nm)^{-1}|\mathbf{H}_\mu|^{-1/2})+O_p(n^{-1}h_{\mu,Z}^{-1}).
\end{eqnarray*}}
\end{lemma}

\begin{lemma}\label{upper_right}
The $1\times 2$ dimensional upper-right block of the matrix \\
$(nm)^{-2}[\mathbf{1},\mathbf{U}_{u},\mathbf{Z}_{z}]^{\top}\mathbf{W}_{\mu,uz}\mathrm{Cov}(\mathbf{Y}|\mathbf{U},\mathbf{Z})\mathbf{W}_{\mu,uz}[\mathbf{1},\mathbf{U}_{u},\mathbf{Z}_{z}]$ is given by
{\small\begin{eqnarray*}
  &&(nm)^{-2}\mathbf{1}^{\top}\mathbf{W}_{\mu,uz}\mathrm{Cov}(\mathbf{Y}|\mathbf{U},\mathbf{Z})\mathbf{W}_{\mu,uz}\left(\begin{matrix}(U_{11}-u,Z_1-z)\\ \vdots\\ (U_{nm}-u,Z_n-z)\end{matrix}\right)\\
  &=&(nm)^{-1}f_{UZ}(u,z)|\mathbf{H}_\mu|^{-1/2}(\mathbf{1}^{\top}\mathbf{H}_\mu^{1/2})R(K_\mu)\left(\gamma(u,u,z)+\sigma^2_\epsilon\right)(1+O_p(tr(\mathbf{H}_\mu^{1/2})))\\
  &+&n^{-1}(f_{UZ}(u,z))^2(\mathbf{1}^{\top}\mathbf{H}_\mu^{1/2})\left[\left(\frac{m-1}{m}\right)h_{\mu,Z}^{-1}R(\kappa)\frac{\gamma(u,u,z)}{f_Z(z)}\right]
  (1+O_p(tr(\mathbf{H}_\mu^{1/2})))\\
  &=&O_p((nm)^{-1}|\mathbf{H}_\mu|^{-1/2}(\mathbf{1}^{\top}\mathbf{H}_\mu^{1/2}))+O_p(n^{-1}(\mathbf{1}^{\top}\mathbf{H}_\mu^{1/2})h_{\mu,Z}^{-1}).
\end{eqnarray*}}

\noindent The $2\times 1$ dimensional lower-left block of the matrix \\
$(nm)^{-2}[\mathbf{1},\mathbf{U}_{u},\mathbf{Z}_{z}]^{\top}\mathbf{W}_{\mu,uz}\mathrm{Cov}(\mathbf{Y}|\mathbf{U},\mathbf{Z})\mathbf{W}_{\mu,uz}[\mathbf{1},\mathbf{U}_{u},\mathbf{Z}_{z}]$\\
is simply the transposed version of this result.
\end{lemma}

\begin{lemma}\label{lower_right}
The $2\times 2$ lower-right block of the matrix \\
$(nm)^{-2}[\mathbf{1},\mathbf{U}_{u},\mathbf{Z}_{z}]^{\top}\mathbf{W}_{\mu,uz}\mathrm{Cov}(\mathbf{Y}|\mathbf{U},\mathbf{Z})\mathbf{W}_{\mu,uz}[\mathbf{1},\mathbf{U}_{u},\mathbf{Z}_{z}]$ is given by
{\small\begin{eqnarray*}
  &&(nm)^{-2}\left(((U_{11}-u),(Z_1-z))^\top,\dots,((U_{nm}-u),(Z_n-z)^\top)\right)\times\\
  &&\times \mathbf{W}_{\mu,uz}\mathrm{Cov}(\mathbf{Y}|\mathbf{U},\mathbf{Z})\mathbf{W}_{\mu,uz}\left(\begin{matrix}(U_{11}-u,Z_1-z)\\ \vdots\\ (U_{nm}-u,Z_n-z)\end{matrix}\right)\\
  &=&(nm)^{-1}f_{UZ}(u,z)|\mathbf{H}_\mu|^{-1/2} \mathbf{H}_\mu R(K_\mu)\left(\gamma(u,u,z)+\sigma^2_\epsilon\right)(1+O_p(tr(\mathbf{H}_\mu^{1/2})))\\
  &+&n^{-1}(f_{UZ}(u,z))^2 \mathbf{H}_\mu\left[\left(\frac{m-1}{m}\right)h_{\mu,Z}^{-1}R(\kappa)\frac{\gamma(u,u,z)}{f_Z(z)}\right]
  (1+O_p(tr(\mathbf{H}_\mu^{1/2})))\\
  &=&O_p((nm)^{-1}|\mathbf{H}_\mu|^{-1/2}\mathbf{H}_\mu)+O_p(n^{-1}\mathbf{H}_\mu h_{\mu,Z}^{-1}).
\end{eqnarray*}}
\end{lemma}

Using the approximations for the bloc-elements of the matrix \\
$(nm)^{-2}[\mathbf{1},\mathbf{U}_{u},\mathbf{Z}_{z}]^{\top}\mathbf{W}_{\mu,uz}\mathrm{Cov}(\mathbf{Y}|\mathbf{U},\mathbf{Z})\mathbf{W}_{\mu,uz}[\mathbf{1},\mathbf{U}_{u},\mathbf{Z}_{z}]$,
given by the Lemmas \ref{upper_left}-\ref{lower_right}, and the
approximation for the matrix
$\left((nm)^{-1}[\mathbf{1},\mathbf{U}_{u},\mathbf{Z}_{z}]^{\top}\mathbf{W}_{\mu,uz}[\mathbf{1},\mathbf{U}_{u},\mathbf{Z}_{z}]\right)^{-1}$, given in \eqref{pre_results}, we can approximate the conditional variance of the bivariate local linear estimator, given in \eqref{Varexpr}. Some tedious yet straightforward matrix algebra leads to $\V(\hat{\mu}(u,z;\mathbf{H}_\mu)|\mathbf{U},\mathbf{Z})=$
{\small
\begin{align*}
&(nm)^{-1}|\mathbf{H}_\mu|^{-1/2}\left\{\frac{R(K_\mu)\left(\gamma(u,u,z)+\sigma^2_\epsilon\right)}{f_{UZ}(u,z)}\right\}\left(1+o_p(1)\right)\\
&+n^{-1}\left[\left(\frac{m-1}{m}\right)h_{\mu,Z}^{-1}R(\kappa)\frac{\gamma(u,u,z)}{f_Z(z)}\right]\left(1+o_p(1)\right),
\end{align*}}
which is asymptotically equivalent to our variance statement of
Theorem \ref{Bias_and_Variance_mu} part \textit{(ii)}.


Next we prove Lemma \ref{upper_left}; the proofs of Lemmas \ref{upper_right} and \ref{lower_right} are equivalent. To show Lemma \ref{upper_left} it will be convenient to split the sum such that\\
$(nm)^{-2}\mathbf{1}^{\top}\mathbf{W}_{\mu,uz}\mathrm{Cov}(\mathbf{Y}|\mathbf{U},\mathbf{Z})\mathbf{W}_{\mu,uz}\mathbf{1}=s_1+s_2$. Using standard procedures from kernel density estimation leads to
{\small\begin{align}
    s_1&=(nm)^{-2}\sum_{ij}(K_{\mu,h}(U_{ij}-u,Z_{i}-z))^2\V(Y_{ij}|\mathbf{U},\mathbf{Z})\label{M_1}\\
    &=(nm)^{-1}|\mathbf{H}_\mu|^{-1/2}f_{UZ}(u,z)R(K_\mu)\left(\gamma(u,u,z)+\sigma^2_\epsilon\right)
    +O((nm)^{-1}|\mathbf{H}_\mu|^{-1/2}\;tr(\mathbf{H}_\mu^{1/2}))\notag,\\
    s_2&=(nm)^{-2}\underset{j\neq
      k}{\sum_{jk}}\sum_{i}h_{\mu,U}^{-1}\kappa(h_{\mu,U}^{-1}(U_{ij}-u))(h_{\mu,Z}^{-1}\kappa(h_{\mu,Z}^{-1}(Z_{i}-z)))^2\,\mathrm{Cov}(Y_{ij},Y_{ik}|\mathbf{U},\mathbf{Z})\times\label{M_2}\\
    &\times h_{\mu,U}^{-1}\kappa(h_{\mu,U}^{-1}(U_{ik}-x))\notag\\
    &=n^{-1}(f_{UZ}(u,z))^2\left[\left(\frac{m-1}{m}\right)h_{\mu,Z}^{-1}R(\kappa)\frac{\gamma(u,u,z)}{f_Z(z)}\right]+O_p(n^{-1}tr(\mathbf{H}_\mu^{1/2}))\notag,
  \end{align}}
Summing up \eqref{M_1}-\eqref{M_2} leads to the result in Lemma \ref{upper_left}. Lemmas \ref{upper_right} and \ref{lower_right} differ from Lemma \ref{upper_left} only with respect to the additional factors $\mathbf{1}^{\top}\mathbf{H}_\mu^{1/2}$ and
$\mathbf{H}_\mu$ which occur due to the usual substitution step for the additional data parts $(U_{ij}-u,Z_i-z)$. 

\subsection{Proof of Theorem \ref{Bias_and_Variance_gamma}}
When neglecting the estimation error in the raw covariances $C_{ijk}$ that is due to estimating the mean function $\mu$, the proof of Theorem \ref{Bias_and_Variance_gamma} follows exactly the same arguments as in the proof of Theorem \ref{Bias_and_Variance_mu} and therefore is omitted. The justification for doing so, follows from the arguments in \cite{jiang2010covariate} (see their proofs of Theorems 3.3 and 3.4.).

\subsection{Proofs of the results in Section \ref{FirstVarTerm}}\label{FD_Suppl}
\subsubsection{Proof of Theorem \ref{AMISE-I_opt_BDW_mu}}
The $\AMISE$ function (i.e., the AMISE function with leading $V_\mu^{I}$ variance term) for the local linear estimator
$\hat{\mu}$ is given by
\begin{align}\label{AMISE.mu.Supl}
  &\AMISE_{\hat{\mu}}\left(h_{\mu,U},h_{\mu,Z}\right)=(nm)^{-1}\,h^{-1}_{\mu,U}\,h^{-1}_{\mu,Z}\,R(K_\mu)\,Q_{\mu,1}+\\
  &+\frac{1}{4}\,\left(\nu_{2}(K_\mu)\right)^2\,\left[h_{\mu,U}^4\,\mathcal{I}_{\mu,UU}+2\,h_{\mu,U}^2\,h_{\mu,Z}^2\,\mathcal{I}_{\mu,UZ}+h_{\mu,Z}^4\,\mathcal{I}_{\mu,ZZ}\right],\notag
\end{align}
\begin{equation*}
  \begin{array}{rcl}
    \text{where \quad\quad}Q_{\mu,1}&=&\int\left(\gamma(u,u,z)+\sigma^2_\epsilon\right)\,d(u,z),\\
    \mathcal{I}_{\mu,UU}&=&\int\left(\mu^{(2,0)}(u,z)\right)^2\,f_{UZ}(u,z)\,d(u,z),\\
    \mathcal{I}_{\mu,ZZ}&=&\int\left(\mu^{(0,2)}(u,z)\right)^2\,f_{UZ}(u,z)\,d(u,z),\quad\text{and}\\
    \mathcal{I}_{\mu,UZ}&=&\int\mu^{(2,0)}(u,z)\mu^{(0,2)}(u,z)\,f_{UZ}(u,z)\,d(u,z).
  \end{array}
\end{equation*}
This is a known expression for the AMISE function of a two-dimensional local linear estimator with a diagonal bandwidth matrix \citep[see, e.g.,][]{herrmann1995bandwidth} and follows from the formulas in \cite{wand1994multivariate}. Minimizing the above AMISE function with respect to $h_{\mu,U}$ and $h_{\mu,Z}$ leads to the optimal bandwidth expressions in Theorem \ref{AMISE-I_opt_BDW_mu} which correspond to the results in \cite{herrmann1995bandwidth}.

It follows directly from Theorem \ref{Bias_and_Variance_mu} that the first variance summand $V_\mu^{I}$ is the leading variance term if the following order relation holds:
\begin{align}\label{eq.order.rel_mu}
  n^{-1}\,h_{\mu,Z}^{-1}&=o\left(n^{-(1+{\theta})}\,h_{\mu,U}^{-1}\,h_{\mu,Z}^{-1}\right),
\end{align}
where we used that by Assumption A-AS $nm\asymp n^{1+\theta}$. Plugging the AMISE optimal bandwidth rates of Theorem \ref{AMISE-I_opt_BDW_mu} into the order relation of Eq.~\eqref{eq.order.rel_mu} leads to the corresponding $\theta$ values of $0\leq \theta < 1/5$ which describe the case we consider here as sparse functional data.

\subsubsection{Proof of Theorem \ref{AMISE-I_opt_BDW_gamma}}
The corresponding $\AMISE$ function (i.e., the AMISE function with leading $V_\gamma^{I}$ variance term) for the local linear estimator $\hat{\gamma}$ is given by
\begin{align}\label{AMISE.gamma}
  &\AMISE_{\hat{\gamma}}\left(h_{\gamma,U},h_{\gamma,Z}\right)=(nM)^{-1}\,h^{-2}_{\gamma,U}\,h^{-1}_{\gamma,Z}\,R(K_\gamma)\,Q_{\gamma,1}+\\
  +\frac{1}{4}\,(\nu_{2}(K_\gamma&))^2\,\left[2\,h_{\gamma,U}^4\,(\mathcal{I}_{\gamma,U_{(1)}U_{(2)}}+\mathcal{I}_{\gamma,U_{(1)}U_{(2)}})+4\,h_{\gamma,U}^2\,h_{\gamma,Z}^2\,\mathcal{I}_{\gamma,U_{(1)}Z}+h_{\gamma,Z}^4\,\mathcal{I}_{\gamma,ZZ}\right],\notag
\end{align}
\begin{equation*}
  \begin{array}{rcl}
\text{where \quad} Q_{\gamma,1}&=&\int\left(\tilde{\gamma}((u_1,u_2),(u_1,u_2),z)+\sigma^2_\varepsilon(u_1,u_2,z)\right)\,d(u_1,u_2,z)\\
    \mathcal{I}_{\gamma,U_{(1)}U_{(1)}}&=&\int\left(\gamma^{(2,0,0)}(u_1,u_2,z)\right)^2\,f_{UUZ}(u_1,u_2,z)\,d(u_1,u_2,z),\\
    \mathcal{I}_{\gamma,U_{(1)}U_{(2)}}&=&\int\left(\gamma^{(2,0,0)}(u_1,u_2,z)\gamma^{(0,2,0)}(u_1,u_2,z)\right)\,f_{UUZ}(u_1,u_2,z)\,d(u_1,u_2,z),\\
    \mathcal{I}_{\gamma,U_{(1)}Z}
    &=&\int\gamma^{(2,0,0)}(u_1,u_2,z)\gamma^{(0,0,2)}(u_1,u_2,z)\,f_{UUZ}(u_1,u_2,z)\,d(u_1,u_2,z),
    \quad\text{and}\\
    \mathcal{I}_{\gamma,ZZ}   &=&\int\left(\gamma^{(0,0,2)}(u_1,u_2,z)\right)^2\,f_{UUZ}(u_1,u_2,z)\,d(u_1,u_2,z).
  \end{array}
\end{equation*}
Equation \eqref{AMISE.gamma} again follows from the
formulas in \cite{wand1994multivariate} and additionally by using the following equalities:\\
$\mathcal{I}_{\gamma,U_{(1)}U_{(1)}}=\mathcal{I}_{\gamma,U_{(2)}U_{(2)}}$,
$\mathcal{I}_{\gamma,U_{(1)}U_{(2)}}=\mathcal{I}_{\gamma,U_{(2)}U_{(1)}}$, and
$\mathcal{I}_{\gamma,U_{(1)}Z}=\mathcal{I}_{\gamma,U_{(2)}Z}$ due to the
symmetry of the covariance function, where the expressions
$\mathcal{I}_{\gamma,U_{(2)}U_{(2)}}$, $\mathcal{I}_{\gamma,U_{(2)}U_{(1)}}$, and
$\mathcal{I}_{\gamma,U_{(2)}Z}$ are defined equivalently to their above defined counterparts.

Minimizing the AMISE function above with respect to $h_{\gamma,U}$ and $h_{\gamma,Z}$ leads to the optimal bandwidth expressions in Theorem \ref{AMISE-I_opt_BDW_gamma}. This is much more cumbersome than for the case of the mean function $\mu$, but can easily done using, e.g., a computer algebra system.

It follows directly from the Theorem \ref{Bias_and_Variance_gamma} that the first variance term $V_\gamma^{I}$ is the leading variance term if the following order relation holds:
\begin{align}\label{eq.order.rel_gamma}
  n^{-1}\,h_{\gamma,Z}^{-1}=o\left(n^{-(1+{2\theta})}\,h_{\gamma,U}^{-2}\,h_{\gamma,Z}^{-1}\right),
\end{align}
where we used that by Assumption A-AS $nM\asymp n^{1+2\theta}$. Plugging the AMISE optimal bandwidth rates of Theorem \ref{AMISE-I_opt_BDW_gamma} into the order relation of Eq.~\eqref{eq.order.rel_gamma} leads to the corresponding $\theta$ values of $0\leq \theta < 1/5$ which describe the case considered here as sparse functional data. Observe that the same $\theta$-threshold value
of $1/5$ applies to both estimators $\hat{\mu}$ and $\hat{\gamma}$.

\subsubsection{Proofs of Corollaries \ref{C_AN_mu} and \ref{C_AN_gamma}}

Corollaries \ref{C_AN_mu} and \ref{C_AN_gamma} follow directly from Theorems \ref{Bias_and_Variance_mu}, \ref{Bias_and_Variance_gamma}, \ref{AMISE-I_opt_BDW_mu}, and \ref{AMISE-I_opt_BDW_gamma} and from applying a standard central limit theorem for iid data.

\subsection{Proofs of the results in Section \ref{SecondVarTerm}}\label{Sec.AMISE.Supl}

\subsubsection{Proof of Theorem \ref{AMISE-II_opt_BDW_mu}}

The AMISE function of $\hat{\mu}$ including both variance terms $V_\mu^{I}$ and $V_\mu^{II}$ is given by
\begin{align}\label{AMISE2.mu}
&\AMISE_{\hat{\mu}}\left(h_{\mu,U},h_{\mu,Z}\right)=
\underbrace{\overbrace{(nm)^{-1}\,h^{-1}_{\mu,U}\,h^{-1}_{\mu,Z}\,R(K_\mu)\,Q_{\mu,1}}^{\int
      V_\mu^{I}(u,z) f_{UZ}(u,z)d(u,z)}}_{\text{2nd
  Order}}
+
\underbrace{\overbrace{n^{-1}\,h^{-1}_{\mu,Z}\,R(\kappa)\,Q_{\mu,2}}^{\int
  V_\mu^{II}(u,z) f_{UZ}(u,z)d(u,z)}}_{\text{1st
  Order}}+\\
&+\frac{1}{4}\,\left(\nu_{2}(K_\mu)\right)^2\,\left[\underbrace{h_{\mu,U}^4\,\mathcal{I}_{\mu,UU}}_{\text{3rd
    Order}}+\underbrace{2\,h_{\mu,U}^2\,h_{\mu,Z}^2\,\mathcal{I}_{\mu,UZ}}_{\text{2nd
Order}}+\underbrace{h_{\mu,Z}^4\,\mathcal{I}_{\mu,ZZ}}_{\text{1st
  Order}}\right],\notag
\end{align}
\begin{equation*}
\begin{array}{rcl}
\text{where \quad}
\mathcal{I}_{\mu,UU}&=&\int(\mu^{(2,0)}(u,z))^2\,f_{UZ}(u,z)\,d(u,z),\\
\mathcal{I}_{\mu,ZZ}&=&\int(\mu^{(0,2)}(u,z))^2\,f_{UZ}(u,z)\,d(u,z),\\
\mathcal{I}_{\mu,UZ}&=&\int \mu^{(2,0)}(u,z)\mu^{(0,2)}(u,z)\,f_{UZ}(u,z)\,d(u,z),\\
Q_{\mu,1}&=&\int\left(\gamma(u,u,z)+\sigma^2_\epsilon\right)\,d(u,z),\quad\text{and}\\
Q_{\mu,2}&=&\int\gamma(u,u,z)\,f_U(u)\,d(u,z).
\end{array}
\end{equation*}

Note that it is impossible to derive explicit AMISE optimal $U$- and $Z$-bandwidth expressions through minimizing Eq.~\eqref{AMISE2.mu} simultaneously for both bandwidths. If the second variance term $V_\mu^{II}$ is the leading variance term, the lowest possible AMISE value can be achieved if there exists a $U$-bandwidth which, first, allows us to profit from the (partial) annulment of the $U$-related bias-variance trade-off, but, second, assures that the second variance term $V_\gamma^{II}$ remains the leading variance term.

The first requirement is achieved if the $U$-bandwidth is of a smaller order of magnitude than the $Z$-bandwidth, i.e., if $h_{\mu,U}=o(h_{\mu,Z})$. This restriction makes those bias components that depend on $h_{\mu,U}$ asymptotically negligible, since it implies that $h_{\mu,U}^2\,h_{\mu,Z}^2=o(h_{\mu,Z}^4)$ and therefore that $h_{\mu,U}^4=o(h_{\mu,U}^2\,h_{\mu,Z}^2)$. The latter two strict inequalities lead to the order relations between the three bias terms as indicated in Eq.~\eqref{AMISE2.mu}. The second requirement is achieved if the $U$-bandwidth does not converge to zero too fast, namely if $mh_{\mu,U}\to\infty$, which implies the order relation between the two variance terms as indicated in Eq.~\eqref{AMISE2.mu}.

Let us initially assume that it is possible to find an $U$-bandwidth that fulfills both the above requirements, namely $h_{\mu,U}=o(h_{\mu,Z})$ and $nh_{\mu,U}\to\infty$. With such an $U$-bandwidth we can make use of the order relations indicated in Eq.~\eqref{AMISE2.mu}. That is, instead of minimizing the $\AMISE$ function in Eq.~\eqref{AMISE2.mu} over both bandwidths, we can minimize the following simpler and asymptotically equivalent AMISE function, which depends only on the $Z$-bandwidth:
\begin{align*}
&\AMISE^{1\text{st Order}}_{\hat{\mu}}\left(h_{\mu,Z}\right)=
n^{-1}\,h^{-1}_{\mu,Z}\,R(\kappa)\,Q_{\mu,2}+
\frac{1}{4}\,\left(\nu_{2}(K_\mu)\right)^2\,h_{\mu,Z}^4\,\mathcal{I}_{\mu,ZZ}.
\end{align*}
The above equation is minimized by the following $Z$-bandwidth:
\begin{align*}
h_{\mu,Z}^{D}&=\left(\frac{R(\kappa)\,Q_{\mu,2}}{n\,\left(\nu_{2}(K_\mu)\right)^2\,\mathcal{I}_{\mu,ZZ}}\right)^{1/5},
\end{align*}
which is that of Eq.~\eqref{h.mu.Z.AMISE2} in Theorem \ref{AMISE-II_opt_BDW_mu}.

We still need to find $U$-bandwidth that fulfills the postulated requirements. To do so we suggest plugging the above optimal $Z$-bandwidth into the $\AMISE$ function in Eq.~\eqref{AMISE2.mu} and minimizing the (then classical) bias-variance trade-off between the asymptotic second order terms, which leads to the following expression for the $U$-bandwidths:
\begin{align*}
h_{\mu,U}^{D}&=\left(\frac{R(K_\mu)\,Q_{\mu,1}}{nm\,\left(\nu_{2}(K_\mu)\right)^2\,\mathcal{I}_{\mu,UZ}}\right)^{1/3}\left(h_{\mu,Z}^{D}\right)^{-1},
\end{align*}
which is that of Eq.~\eqref{h.mu.U.AMISE2} in Theorem \ref{AMISE-II_opt_BDW_mu}.

In order to check whether this $U$-bandwidth actually fulfills the two necessary requirements, we apply some rearrangements. Using that by Assumption AS $m\asymp n^\theta$, leads to the following more transparent presentation of the bandwidth rates:
\begin{align}
h_{\mu,Z}^{D}\asymp m^{-1/(5\,\theta)}\quad\text{and}\quad h_{\mu,U}^{D}\asymp m^{-\eta_\mu(\theta)}&\quad\text{with}\quad
\eta_\mu(\theta)=\frac{1}{3}+\frac{2}{15\,\theta}\label{h.mu.U.AMISE.rate}
\end{align}
With Eq.~\eqref{h.mu.U.AMISE.rate} it is easily verified that the necessary requirements ($h_{\mu,U,\AMISE}=o(h_{\mu,Z}^{D})$ and $mh_{\mu,U}^{D}\to\infty$) are fulfilled iff $\theta>1/5$.

\subsubsection{Proof of Theorem \ref{AMISE-II_opt_BDW_gamma}}
The $\AMISE$ expression of $\hat{\gamma}$ including both variance terms $V_\gamma^{I}$ and $V_\gamma^{II}$  is given by
\begin{align}\label{AMISE2.gamma}
&\AMISE_{\hat{\gamma}}\left(h_{\gamma,U},h_{\gamma,Z}\right)=\overbrace{(nM)^{-1}\,h^{-2}_{\gamma,U}\,h^{-1}_{\gamma,Z}\,R(K_\gamma)\,Q_{\gamma,1}}^{\text{2nd
  Order}}+
\overbrace{n^{-1}\,h^{-1}_{\gamma,Z}\,R(\kappa)\,Q_{\gamma,2}}^{\text{1st
  Order}}+\\ +\frac{1}{4}\,(\nu_{2}(K_\gamma&))^2\,\left[\underbrace{2\,h_{\gamma,U}^4\,(\mathcal{I}_{\gamma,U_{(1)}U_{(1)}}+\mathcal{I}_{U_{(1)}U_{(2)}})}_{\text{3rd
    Order}}+\underbrace{4\,h_{\gamma,U}^2\,h_{\gamma,Z}^2\,\mathcal{I}_{\gamma,U_{(1)}Z}}_{\text{2nd
Order}}+\underbrace{h_{\gamma,Z}^4\,\mathcal{I}_{\gamma,ZZ}}_{\text{1st
  Order}}\right],\notag
\end{align}
\begin{equation*}
\begin{array}{rcl}
\text{where \quad}
\mathcal{I}_{\gamma,U_{(1)}U_{(1)}}&=&\int\left(\gamma^{(2,0,0)}(u_1,u_2,z)\right)^2\,f_{UUZ}(u_1,u_2,z)\,d(u_1,u_2,z),\\
\mathcal{I}_{\gamma,U_{(1)}U_{(2)}}&=&\int\left(\gamma^{(2,0,0)}(u_1,u_2,z)\gamma^{(0,2,0)}(u_1,u_2,z)\right)\,f_{UUZ}(u_1,u_2,z)\,d(u_1,u_2,z),\\
\mathcal{I}_{\gamma,U_{(1)}Z}
    &=&\int\gamma^{(2,0,0)}(u_1,u_2,z)\gamma^{(0,0,2)}(u_1,u_2,z)\,f_{UUZ}(u_1,u_2,z)\,d(u_1,u_2,z),\\
\mathcal{I}_{\gamma,ZZ}   &=&\int\left(\gamma^{(0,0,2)}(u_1,u_2,z)\right)^2\,f_{UUZ}(u_1,u_2,z)\,d(u_1,u_2,z),\\
Q_{\gamma,1}&=&\int\left(\tilde{\gamma}((u_1,u_2),(u_1,u_2),z)+\sigma^2_\varepsilon(u_1,u_2,z)\right)\,d(u_1,u_2,z),\quad\text{and}\\
Q_{\gamma,2}&=&\int\tilde{\gamma}((u_1,u_2),(u_1,u_2),z)\,f_{UU}(u_1,u_2)\,d(u_1,u_2,z)
\end{array}
\end{equation*}

By the same reasoning as in the preceding section, we initially determine requirements on the $U$-bandwidth that maintain the order relation between the two variance terms as indicated in Eq.~\eqref{AMISE2.gamma}. The first requirement is that $h_{\gamma,U}=o(h_{\gamma,Z})$. This restriction makes those bias components that depend on $h_{\gamma,U}$ asymptotically negligible, since it implies that $h_{\gamma,U}^2\,h_{\gamma,Z}^2=o(h_{\gamma,Z}^4)$ and therefore that $h_{\gamma,U}^4=o(h_{\gamma,U}^2\,h_{\gamma,Z}^2)$. The latter leads to the order relations between the three bias terms as indicated in Eq.~\eqref{AMISE2.gamma}. The second requirement is that the $U$-bandwidth does not converge to zero too fast, namely that $Mh_{\gamma,U}^2\to\infty$, which implies the order relation between the first two variance terms as indicated in Eq.~\eqref{AMISE2.gamma}.

Under these requirements on the $U$-bandwidths, we can minimize the following simpler and asymptotically equivalent AMISE function, which depends only on the $Z$-bandwidth:
\begin{align*}
&\AMISE^{1\text{st Order}}_{\hat{\gamma}}\left(h_{\gamma,Z}\right)=n^{-1}\,h^{-1}_{\gamma,Z}\,R(\kappa)\,Q_{\gamma,2}+ \frac{1}{4}\,(\nu_{2}(K_\gamma))^2\,h_{\gamma,Z}^4\,\mathcal{I}_{\gamma,ZZ}.
\end{align*}
The above equation is minimized by the following $Z$-bandwidth
\begin{align*}
h_{\gamma,Z}^{D}&=\left(\frac{R(\kappa)\,Q_{\gamma,2}}{n\,\left(\nu_{2}(K_\gamma)\right)^2\,\mathcal{I}_{\gamma,ZZ}}\right)^{1/5},
\end{align*}
which is that of Eq.~\eqref{h.gamma.Z.AMISE2} in Theorem \ref{AMISE-II_opt_BDW_gamma}.

Parallel to the preceding section, we determine the $U$-bandwidth by plugging the above optimal $Z$-bandwidth into the $\AMISE$ function in Eq.~\eqref{AMISE2.gamma} and by minimizing the (then classical) bias-variance trade-off between the asymptotic second order terms, which leads to the following expression for the $U$-bandwidths:
\begin{align*}
h_{\gamma,U}^{D}&=\left(\frac{R(K_\gamma)\,Q_{\gamma,1}}{nM\,\left(\nu_{2}(K_\gamma)\right)^2\,\mathcal{I}_{\gamma,U_{(1)}Z}}\right)^{1/4}\left(h_{\gamma,Z}^{D}\right)^{-3/4},
\end{align*}
which is that of Eq.~\eqref{h.gamma.U.AMISE2} in Theorem \ref{AMISE-II_opt_BDW_gamma}.

In order to check whether this $U$-bandwidth actually fulfills the two necessary requirements, we apply some rearrangements. Using that by Assumption AS $m\asymp n^\theta$ and that by construction $M\asymp m^2$, leads to the following more transparent presentation of the bandwidth rates:
\begin{align}
h_{\gamma,Z}^{D}\asymp M^{-1/(10\,\theta)}
\quad\text{and}\quad h_{\gamma,U}^{D}\asymp M^{-\eta_\gamma(\theta)}&\quad\text{with}\quad
\eta_\gamma(\theta)=\frac{1}{4}+\frac{1}{20\,\theta}.\label{h.gamma.U.AMISE.rate}
\end{align}
With Eq.~\eqref{h.gamma.U.AMISE.rate} it is easily verified that the necessary requirements, i.e., that $h_{\gamma,U,\AMISE}=o(h_{\gamma,Z,\AMISE})$ and $Mh_{\gamma,U,\AMISE}^2\to\infty$, are fulfilled iff $\theta>1/5$.

\subsubsection{Proofs of Corollaries \ref{C_AN_mu_2} and \ref{C_AN_gamma_2}}

Corollaries \ref{C_AN_mu_2} and \ref{C_AN_gamma_2} follow directly from Theorems \ref{Bias_and_Variance_mu} and
\ref{Bias_and_Variance_gamma} and from applying a standard central limit theorem for iid data.

\bibliographystyle{Chicago}
\bibliography{bibfile}

\end{document}